\documentclass[superscriptaddress,prc,twocolumn,preprintnumbers,amsmath,amssymb,bibnotes,altaffilletter]{revtex4}
\usepackage{graphicx}
\usepackage{xspace}
\usepackage{sidecap}
\usepackage{amsmath}
\usepackage{amssymb}
\usepackage{url}
\usepackage{braket} 
\usepackage{slantsc}
\usepackage[symbol]{footmisc}
\usepackage[T1]{fontenc}

\def\nuc#1#2{${}^{#1}$#2}
\def\mee{$\langle m_{\beta\beta} \rangle$}

\def\BBz{$\beta\beta(0\nu)$}

\def\Mz{$M_{0\nu}$}

\def\be{\begin{equation}}
\def\ee{\end{equation}}
\def\cpRty{cnts/(ROI-t-y)}
\def\onecpRty{1~cnt/(ROI-t-y)}
\def\threecpRty{3~cnts/(ROI-t-y)}
\def\ppc{P-PC}                          

\def\MJ{{\sc Majorana}}             
\def\DEM{{\sc Demonstrator}}             
\def\MJDEMbf{\bfseries{\scshape{Majorana Demonstrator}}}

\def\MJDEMit{\itshape{\scshape{Majorana Demonstrator}}}
\newcommand{\Gerda}{GERDA}
\newcommand{\GF}{\textsc{Geant4}}
\newcommand{\MaGe}{\textsc{MaGe}}

\begin{document}
\title{The \MJDEMbf\ Neutrinoless Double-Beta Decay Experiment}

\newcommand{\alberta}{Centre for Particle Physics, University of Alberta, Edmonton, AB, Canada}
\newcommand{\blhill}{Department of Physics, Black Hills State University, Spearfish, SD, USA}
\newcommand{\ITEP}{Institute for Theoretical and Experimental Physics, Moscow, Russia}
\newcommand{\JINR}{Joint Institute for Nuclear Research, Dubna, Russia}
\newcommand{\lbnl}{Nuclear Science Division, Lawrence Berkeley National Laboratory, Berkeley, CA, USA}
\newcommand{\lanl}{Los Alamos National Laboratory, Los Alamos, NM, USA}
\newcommand{\uw}{Center for Experimental Nuclear Physics and Astrophysics, 
and Department of Physics, University of Washington, Seattle, WA, USA}
\newcommand{\uchic}{Department of Physics, University of Chicago, Chicago, IL, USA}
\newcommand{\unc}{Department of Physics and Astronomy, University of North Carolina, Chapel Hill, NC, USA}
\newcommand{\duke}{Department of Physics, Duke University, Durham, NC, USA}
\newcommand{\ncsu}{Department of Physics, North Carolina State University, Raleigh, NC, USA}
\newcommand{\ornl}{Oak Ridge National Laboratory, Oak Ridge, TN, USA}
\newcommand{\ou}{Research Center for Nuclear Physics and Department of Physics, Osaka University, Ibaraki, Osaka, Japan}
\newcommand{\pnnl}{Pacific Northwest National Laboratory, Richland, WA, USA}
\newcommand{\ttu}{Tennessee Tech University, Cookeville, TN, USA}
\newcommand{\sdsmt}{South Dakota School of Mines and Technology, Rapid City, SD, USA}
\newcommand{\sjtu}{Shanghai Jiaotong University, Shanghai, China}
\newcommand{\usc}{Department of Physics and Astronomy, University of South Carolina, Columbia, SC, USA}
\newcommand{\usd}{Department of Physics, University of South Dakota, Vermillion, SD, USA} 
\newcommand{\ut}{Department of Physics and Astronomy, University of Tennessee, Knoxville, TN, USA}
\newcommand{\tunl}{Triangle Universities Nuclear Laboratory, Durham, NC, USA}

\affiliation{\lbnl} 
\affiliation{\pnnl} 
\affiliation{\usc} 
\affiliation{\ornl}  
\affiliation{\ITEP} 
\affiliation{\lanl} 
\affiliation{\JINR}
\affiliation{\tunl} 
\affiliation{\duke}  
\affiliation{\sdsmt}  
\affiliation{\ncsu}
\affiliation{\uw} 
\affiliation{\ut} 
\affiliation{\ou} 
\affiliation{\unc}
\affiliation{\usd}
\affiliation{\blhill}
\affiliation{\ttu} 
\affiliation{\sjtu}  

\author{N.~Abgrall}\affiliation{\lbnl}		
\author{E.~Aguayo}\affiliation{\pnnl} 
\author{F.T.~Avignone~III}\affiliation{\usc}\affiliation{\ornl}
\author{A.S.~Barabash}\affiliation{\ITEP}
\author{F.E.~Bertrand}\affiliation{\ornl}
\author{M.~Boswell}\affiliation{\lanl} 
\author{V.~Brudanin}\affiliation{\JINR}
\author{M.~Busch}\affiliation{\duke}\affiliation{\tunl}	
\author{A.S.~Caldwell}\affiliation{\sdsmt}	
\author{Y-D.~Chan}\affiliation{\lbnl}
\author{C.D.~Christofferson}\affiliation{\sdsmt} 
\author{D.C.~Combs}\affiliation{\ncsu}\affiliation{\tunl}  
\author{J.A.~Detwiler}\affiliation{\uw}	
\author{P.J.~Doe}\affiliation{\uw} 
\author{Yu.~Efremenko}\affiliation{\ut}
\author{V.~Egorov}\affiliation{\JINR}
\author{H.~Ejiri}\affiliation{\ou}
\author{S.R.~Elliott}\thanks{Corresponding Author}\email{elliotts@lanl.gov}\affiliation{\lanl}
\author{J.~Esterline}\affiliation{\duke}\affiliation{\tunl}
\author{J.E.~Fast}\affiliation{\pnnl}
\author{P.~Finnerty}\affiliation{\unc}\affiliation{\tunl}  
\author{F.M.~Fraenkle}\affiliation{\unc}\affiliation{\tunl} 
\author{A.~Galindo-Uribarri}\affiliation{\ornl}	
\author{G.K.~Giovanetti}\affiliation{\unc}\affiliation{\tunl}  
\author{J. Goett}\affiliation{\lanl}	
\author{M.P.~Green}\affiliation{\unc}\affiliation{\tunl}  
\author{J.~Gruszko}\affiliation{\uw}		
\author{V.E.~Guiseppe}\affiliation{\usd}	
\author{K.~Gusey}\affiliation{\JINR}
\author{A.L.~Hallin}\affiliation{\alberta}
\author{R.~Hazama}\affiliation{\ou}
\author{A.~Hegai}\altaffiliation{Permanent address: Tuebingen University, Tuebingen, Germany}\affiliation{\lbnl}  
\author{R.~Henning}\affiliation{\unc}\affiliation{\tunl}
\author{E.W.~Hoppe}\affiliation{\pnnl}
\author{S.~Howard}\affiliation{\sdsmt}  
\author{M.A.~Howe}\affiliation{\unc}\affiliation{\tunl}
\author{K.J.~Keeter}\affiliation{\blhill}
\author{M.F.~Kidd}\affiliation{\ttu}	
\author{A.~Knecht}\affiliation{\uw}	
\author{O.~Kochetov}\affiliation{\JINR}
\author{S.I.~Konovalov}\affiliation{\ITEP}
\author{R.T.~Kouzes}\affiliation{\pnnl}
\author{B.D.~LaFerriere}\affiliation{\pnnl}   
\author{J.~Leon}\affiliation{\uw}	
\author{L.E.~Leviner}\affiliation{\ncsu}\affiliation{\tunl}
\author{J.C.~Loach}\affiliation{\sjtu}\affiliation{\lbnl}	
\author{S.~MacMullin}\affiliation{\unc}\affiliation{\tunl} 
\author{R.D.~Martin}\affiliation{\lbnl}	
\author{S.~Mertens}\affiliation{\lbnl}	
\author{L.~Mizouni}\affiliation{\usc}	\affiliation{\pnnl}
\author{M.~Nomachi}\affiliation{\ou}
\author{J.L.~Orrell}\affiliation{\pnnl}
\author{C. O'Shaughnessy}\affiliation{\unc}\affiliation{\tunl}	
\author{N.R.~Overman}\affiliation{\pnnl}  
\author{D.G.~Phillips II}\affiliation{\ncsu}\affiliation{\tunl}  
\author{A.W.P.~Poon}\affiliation{\lbnl}
\author{K.~Pushkin}\affiliation{\usd} 
\author{D.C.~Radford}\affiliation{\ornl}
\author{K.~Rielage}\affiliation{\lanl}
\author{R.G.H.~Robertson}\affiliation{\uw}
\author{M.C.~Ronquest}\affiliation{\lanl}	
\author{A.G.~Schubert}\affiliation{\uw}		
\author{B.~Shanks}\affiliation{\unc}\affiliation{\tunl}	
\author{T.~Shima}\affiliation{\ou}
\author{M.~Shirchenko}\affiliation{\JINR}
\author{K.J.~Snavely}\affiliation{\unc}\affiliation{\tunl}	
\author{N.~Snyder}\affiliation{\usd}	
\author{D. Steele}\affiliation{\lanl}
\author{J.~Strain}\affiliation{\unc}\affiliation{\tunl}
\author{A.M.~Suriano}\affiliation{\sdsmt} 
\author{J.~Thompson}\affiliation{\blhill} 
\author{V.~Timkin}\affiliation{\JINR}
\author{W.~Tornow}\affiliation{\duke}\affiliation{\tunl}
\author{R.L.~Varner}\affiliation{\ornl}  
\author{S.~Vasilyev}\affiliation{\ut}	
\author{K.~Vetter}\altaffiliation{Alternate address: Department of Nuclear Engineering, University of California, Berkeley, CA, USA}\affiliation{\lbnl}%
\author{K.~Vorren}\affiliation{\unc}\affiliation{\tunl} 
\author{B.R.~White}\affiliation{\ornl}	
\author{J.F.~Wilkerson}\affiliation{\unc}\affiliation{\tunl}\affiliation{\ornl}    
\author{W.~Xu}\affiliation{\lanl}  
\author{E.~Yakushev}\affiliation{\JINR}
\author{A.R.~Young}\affiliation{\ncsu}\affiliation{\tunl}
\author{C.-H.~Yu}\affiliation{\ornl}
\author{V.~Yumatov}\affiliation{\ITEP}
			
\collaboration{{\sc{Majorana}} Collaboration}
\noaffiliation

\begin{abstract}
  The \MJ\ \DEM\ will search for the neutrinoless double-beta (\BBz) decay of the isotope \nuc{76}{Ge} with a mixed array of enriched and natural germanium detectors.  The observation of this rare decay would indicate the neutrino is its own antiparticle, demonstrate that lepton number is not conserved, and provide information on the absolute mass scale of the neutrino. The \DEM\ is being assembled at the 4850-foot level of the Sanford Underground Research Facility in Lead, South Dakota. The array will be situated in a low-background environment and surrounded by passive and active shielding. Here we describe the science goals of the \DEM\ and the details of its design.

\end{abstract}
\maketitle

\section{Introduction}
\subsection{Neutrinoless Double-beta decay}

Despite being discovered well over a decade ago~\cite{Ash04,Ahm04,Abe08}, the incorporation of neutrino mass and mixing into the Standard Model (SM) of particle physics remains elusive. A minimalistic Higgs coupling leaves the SM fine-tuned, with the neutrino masses lying some 6 orders-of-magnitude or more below that of the other SM leptons. Avoiding such unnaturalness requires new physics. One highly attractive option, afforded by the electric neutrality of the neutrino, is the addition of a lepton-number-violating Majorana mass term~\cite{Majorana1937}. Majorana neutrinos have the novel property that particle and antiparticle are distinguished only by chirality. A Majorana mass term provides a natural explanation for the lightness of the SM neutrino via the seesaw mechanism~\cite{yan79,gel79}. Majorana neutrinos also provide plausible scenarios for leptogenesis capable of accounting for the excess of matter over antimatter in the observable universe~\cite{fuk86,DiBari2012}. 

Neutrinoless double-beta (\BBz) decay searches represent the only viable experimental method for testing the Majorana nature of neutrinos ~\cite{Zralek1997}. The observation of this process would immediately imply that lepton number is violated and that neutrinos are Majorana particles~\cite{sch82}. The decay rate may be written as
\begin{equation}
\label{eqn:BBhalf}
(T_{1/2}^{0\nu})^{-1} = G^{0\nu}\left|\mbox{\Mz}\right|^2\left(\frac{\mbox{\mee}}{m_e}\right)^2,
\end{equation}
where $G^{0\nu}$ is a phase space factor including the couplings, \Mz\ is a nuclear matrix element, $m_e$ is the electron mass, and \mee\ is the effective Majorana neutrino mass. The latter is given by
\begin{equation}
\mbox{\mee} = \left| \sum_{i=1}^3 U_{ei}^2 m_i \right|
\end{equation}
where $U_{ei}$ specifies the admixture of neutrino mass eigenstate $i$ in the electron neutrino. 

Until very recently, the most sensitive limits on \BBz\ decay came from Ge detectors enriched in \nuc{76}{Ge}, namely the Heidelberg-Moscow experiment~\cite{bau99}, and the IGEX experiment~\cite{Aalseth1999,Aal02a,aal04}.  Recent results from the EXO-200 experiment~\cite{Ack11,Auger2012} and from the KamLAND-Zen experiment~\cite{Gando2012,Gando2013} have claimed stronger bounds on neutrino mass. It is difficult, however, to determine the best limit because of the uncertainties in the nuclear matrix elements. Previous-generation \nuc{76}{Ge} experiments also yielded a claim of the direct observation of \BBz\ decay by Klapdor-Kleingrothaus et al.~\cite{kla06}. This  claim has not been widely accepted by the neutrino community~\cite{aal02b,zde02a,fer02}. While the EXO-200 and KamLAND-Zen results are in conflict with this claim, the recent results from GERDA~\cite{Agostini2013a,Agostini2013b,Agostini2013c} show that the observed peak is not an indication of \BBz\ decay in a \nuc{76}{Ge} experiment, which can compare its results with the Klapdor-Kleingrothaus claim without depending on the nuclear matrix elements. For recent comprehensive experimental and theoretical reviews, see Refs.~\cite{ell02, ell04, bar04, Eji05, avi05, avi08, bar11, Rode11, Elliott2012, Vergados2012}.

A measurement of the \BBz\ decay rate would yield information on the absolute neutrino mass. Measurements of atmospheric, solar, and reactor neutrino oscillation~\cite{Beringer2012} indicate a large parameter space for discovery of \BBz\ decay just beyond current experimental bounds below \mee\ $\sim$50 meV. Moreover, evidence from the SNO experiment~\cite{Ahm04} of a clear departure from non-maximal mixing in solar neutrino oscillation, implies a minimum effective Majorana neutrino mass of $\sim$15 meV for the inverted hierarchy scenario. This target is within reach of next-generation \BBz\ searches. An experiment capable of observing this minimum rate would therefore definitively determine the Majorana or Dirac nature of the neutrino for inverted-hierarchical neutrino masses.

Recent developments in germanium detector technology make a \BBz\ decay search feasible using \nuc{76}{Ge}. In this article we describe the \MJ\ \DEM, an experimental effort under construction in the Sanford Underground Research Facility (SURF) whose goal is to demonstrate the techniques required for a definitive next-generation \BBz\ decay experiment with enriched Ge detectors. The {\sc Demonstrator} will also test the Klapdor-Kleingrothaus claim, and will be sensitive to other non-\BBz\ physics signals in Ge. A complementary effort in Ge with similar sensitivity, the GERDA experiment~\cite{Ackermann2013}, is presently operating in the Laboratori Nazionali del Gran Sasso (LNGS). The GERDA and \MJ\ collaborations intend to join in a proposal for the construction of a tonne-scale experiment. A nearly background-free tonne-scale \nuc{76}{Ge} experiment would be sensitive to effective Majorana neutrino masses below $\sim$20~meV, potentially covering the parameter space corresponding to the inverted neutrino-mass hierarchy.

\subsection{Non \BBz\ Physics with the \MJDEMbf}

The Ge-detector design used by \MJ\ has an energy threshold of $\sim$500 eV. This low threshold not only is critical for reducing \BBz\ background (Sec.~\ref{Sec:Background}), but in combination with low backgrounds, opens up new physics programs for the \MJ\ \DEM. Recent experiments~\cite{Aal08, Aalseth2011, Aalseth2011c, Aalseth2011b} have shown the sensitivity of P-type, Point-Contact (\ppc) Ge detectors to light WIMP ($<10~\mathrm{GeV}/c^2$) dark matter via direct detection. A very recent excess of low energy events reported by the CDMS collaboration~\cite{Agnese2013} lends further motivation for doing such a measurement. The \DEM\ may improve the current light WIMP limits by two orders of magnitude~\cite{Giovanetti2012}. 

In addition to light WIMPS, \MJ\ will also be sensitive to solar axions that interact in the Ge crystals via several possible axion-electron coupling mechanisms. One of these mechanisms of particular interest to \MJ\ relies on the Primakoff conversion of axions into photons within the Ge crystal lattice when a Bragg condition is satisfied~\cite{Avignone1998,Creswick1998}. This technique requires knowledge of the crystal axis orientation relative to the Sun at all times to maximize sensitivity. The collaboration will measure the detector crystal orientation for this purpose. The \MJ\ \DEM\ can also search for solar axions generated by the bremsstrahlung mechanism in the sun~\cite{Derbin2011}, and detected by the axioelectric effect~\cite{Avignone2009}. Since this axion spectrum peaks at about 0.6 keV and falls sharply by an order of magnitude by about 3 keV, the low threshold and background are key for this measurement.

\MJ\ will also be sensitive to Pauli Exclusion Principle Violating (PEPV) decays~\cite{Elliott2012a}. In this process, an atomic electron in a Ge atom spontaneously transitions from an upper shell to the K shell, resulting in 3 ground shell electrons. During this de-excitation a photon of energy close to that of a K-shell X-ray (10 keV) is emitted. The slight difference in energy is due to the extra screening of the nucleus by the two K-shell electrons. The detection of this photon would indicate a PEPV decay. Given the large number of atoms present in 40~kg of Ge, this will be a sensitive test of PEPV effects.  

\ppc\ detectors were originally proposed for detecting coherent nuclear scattering of reactor neutrinos~\cite{Barbeau2007} and there is interest in using HPGe detectors to do a similar measurement with higher energy neutrinos at the Spallation Neutron Source at Oak Ridge National Laboratory or similar sources~\cite{Scholberg2006, Anderson2008}. A cryostat full of natural germanium detectors, similar to that planned for the \DEM, deployed at a shallow underground site near the Spallation Neutron Source target should have sufficient sensitivity to make an observation this process. Such an effort could demonstrate the feasibility of \ppc\ technology for reactor monitoring and nuclear treaty verification.

\section{The \MJDEMbf, An Overview}
The \MJ\ \DEM\ is an array of enriched and natural germanium detectors that will search for the \BBz\ decay of the isotope \nuc{76}{Ge}. The specific goals of the \MJ\ \DEM\ are:

\begin{enumerate}
\item Demonstrate a path forward to achieving a background rate at or below \onecpRty\ in the 4-keV region of interest (ROI) around the  2039-keV  Q-value  for \nuc{76}{Ge} \BBz-decay. This is required for tonne-scale germanium-based searches that will probe the inverted-hierarchy parameter space for \BBz\ decay.
\item Show technical and engineering scalability toward a tonne-scale instrument.
\item Test the Klapdor-Kleingrothaus et al. claim~\cite{kla06}. 
\item Perform searches for physics beyond the standard model, such as the search for dark matter and axions.
\end{enumerate}

\MJ\ utilizes the demonstrated benefits of enriched high-purity germanium (HPGe) detectors. These include intrinsically low-background source material, understood enrichment chemistry, excellent energy resolution, and sophisticated event reconstruction. The main technical challenge is the reduction of environmental ionizing radiation backgrounds by about a factor 100 below what has been achieved in previous experiments. 

We have designed a modular instrument composed of two cryostats built from ultra-pure electroformed copper, with each cryostat capable of housing over 20 kg of \ppc\ detectors. \ppc\ detectors were chosen after extensive R\&D by the collaboration and each  has a mass of about 0.6-1.0 kg. The baseline plan calls for 30 kg of the detectors to be built from Ge material enriched to 86\% in isotope 76 and 10 kg fabricated from natural-Ge (7.8\% \nuc{76}{Ge}). The modular approach will allow us to assemble and optimize each cryostat independently, providing a fast deployment with minimum interference on already-operational detectors. 

Starting from the innermost cavity, the cryostats will be surrounded by an inner layer of electroformed copper, an outer layer of Oxygen-Free High thermal Conductivity (OFHC) copper, high-purity lead, an active muon veto, polyethylene, and borated polyethylene. The cryostats, copper, and lead shielding will all be enclosed in a radon exclusion box. The entire experiment will be located in a clean room at the 4850' level (1478 m) of the Sanford Underground Research Facility (SURF) in Lead, South Dakota.

\section{The \ppc-Detector Technology}
\label{Sec:PPCTech}
At the heart of \MJ\ is its enriched, p-type point-contact HPGe detectors~\cite{luk89,Barbeau2007}. These detectors have all the benefits of coax HPGE detectors traditionally used for \BBz, but also possess superb pulse shape analysis (PSA) discrimination between single-site interactions (such as \BBz-decay events) and multi-site interaction events (such as Compton scattering of $\gamma$-ray backgrounds), making them highly suitable for \BBz\ searches. Their small capacitance results in superb energy resolution and a low energy threshold, making them suitable for event correlation techniques using x rays. Furthermore, they are relatively robust and simple to produce. Their simplicity has the advantage of reducing the characterization studies and detector-to-detector tuning required for an effective PSA algorithm.

Like coaxial Ge detectors, \ppc\ detectors are cylindrical in shape. The electron holes, however, are collected on a small, shallow contact, rather than a long extended electrode as in a coaxial detector. In the detectors made for \MJ, this ``point contact'' varies in diameter from about 2 to 6.5 mm, and in depth from less than a micron (for implanted contacts) to a few mm. Since they have no long inner contact, \ppc\ detectors are generally limited in their length-to-diameter aspect ratio. If a crystal is too long, it can result in having a ``pinch-off'' island of undepleted material in the center, especially if the point-contact end of the crystal has a lower net impurity concentration than the opposite end. This can be alleviated by using crystals with larger impurity gradients, and ensuring that the point contact is placed at the end of the crystal where the material is of higher impurity (generally the seed end).

The \MJ\ collaboration has procured 20 kg of natural-germanium modified-BEGe detectors from CANBERRA Industries~\cite{BEGe}, modified so as to not have the thin front window that permits sensitivity to low-energy external $\gamma$ rays. These detectors typically have masses in the range of 600-700 g, and use a thin, implanted contact. 
Detectors from enriched \nuc{76}{Ge} material are being produced by AMETEK/ORTEC~\cite{ORTEC}. These detectors have a mass of around 1 kg each, with a greater length-to-diameter ratio than the BEGe detectors. We anticipate that approximately 30 kg of these detectors will be produced for the \MJ\ \DEM\ from the 41.6 kg of 86\%-enriched \nuc{76}{Ge} material supplied to AMETEK/ORTEC.

\begin{figure}[ht]
\begin{center}
\includegraphics[width=8cm]{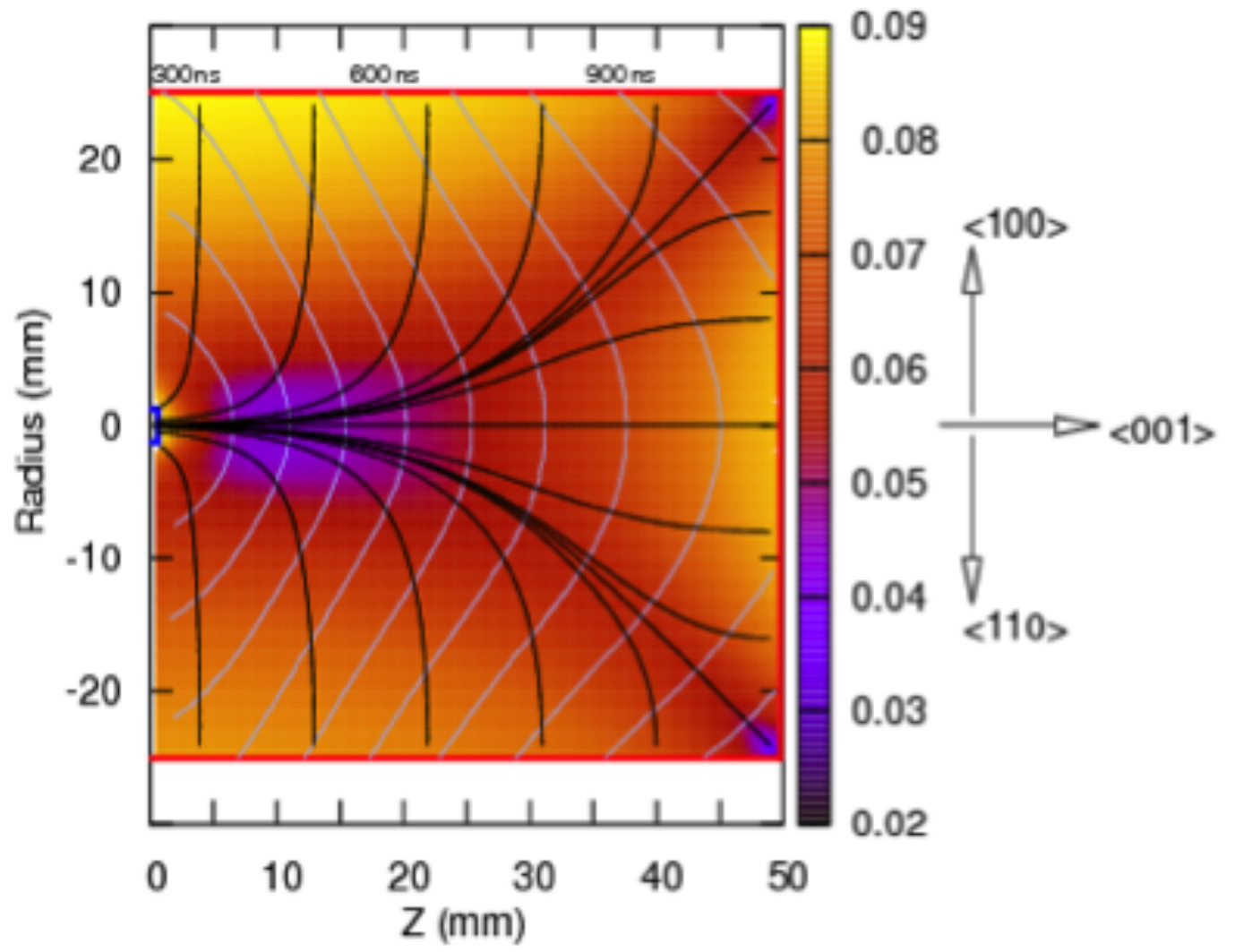}
\caption{Calculated hole drift in a 5-cm diameter $\times$ 5-cm long \ppc\ detector.}
\label{fig:DriftCalc}
\end{center}
\end{figure}

Figure~\ref{fig:DriftCalc} illustrates our modeling of a sample \ppc\ detector, 5 cm in diameter and 5 cm long. The color scale shows hole drift speeds, in mm/ns, the black lines show charge drift trajectories, and the light grey lines show ``isochrones'' - loci of equal hole drift time for events in the detector bulk. We have adapted such drift-time calculations to create a PSA heuristic for \GF\ simulations (see Sec.~\ref{Sec:Simulations}.) of the remaining background in the \MJ\ \DEM\ following the PSA cut. For this heuristic, multiple interactions within an event are examined for their relative drift times, and that information is combined with each individual energy deposit to determine whether the PSA algorithm would be capable of rejecting the event.

\begin{figure}[ht]
\begin{center}
\includegraphics[width=10 cm]{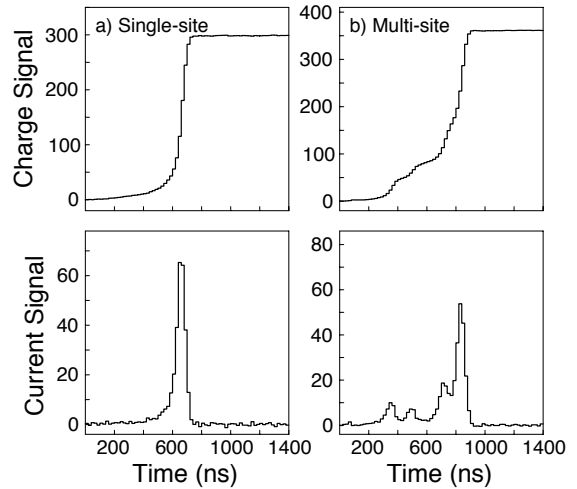}
\caption{Current and charge pulse response of a P-PC detector to single- and multi-site $\gamma$-ray events. The pulse shapes in (a) show the charge (top) and current (bottom) signals resulting from a typical single-site interaction, while (b) shows how the pulse-shape response to a multi-site interaction is clearly different.}
\label{fig:ChargeCurrentPulse}
\end{center}
\end{figure}

Measured signals from a \ppc\ detector are shown in Fig.~\ref{fig:ChargeCurrentPulse}. Both current and charge pulses are shown, for both a single-site (a) and a multi-site (b) $\gamma$-ray event. The difference in signal shape is readily apparent, with four distinct interactions evident in (b). The \MJ\ collaboration uses two different types of PSA algorithm to discriminate between these two classes of events. The first of these, developed by the GERDA collaboration~\cite{dusa09}, compares the maximum height of the current pulse (A) to the total energy of the event (E) as determined from the height of the charge pulse. Multiple interactions result in multiple charge pulses separated in time, and therefore in a reduced value of A/E.

\begin{figure*}[ht]
\begin{center}
\includegraphics[width=16cm]{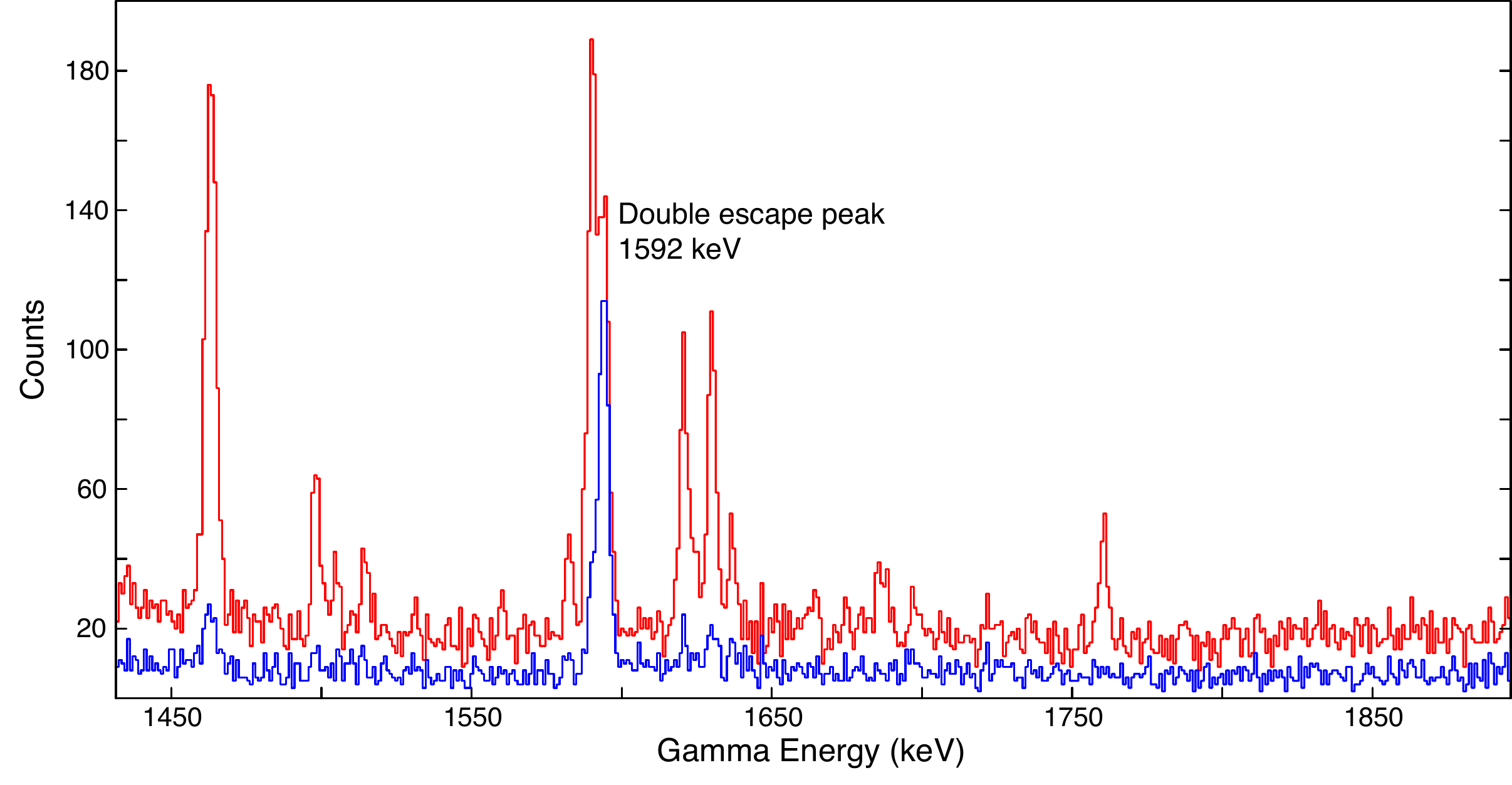}
\caption{Pulse-shape analysis results for \ppc\ data.
The red spectrum is for all events within the energy range, while the blue spectrum is for events that pass the PSA cut.}
\label{fig:PSAspectrum}
\end{center}
\end{figure*}

An alternative approach~\cite{Cooper2011} uses a library of unique, measured single-site signals to perform event-by-event $\chi^2$ fitting of experimental pulse shapes. A method for building this library from a large number of measured signals has been developed and tested with simulation and experimental studies. Results of this optimized PSA algorithm on \ppc\ data are shown in Fig.~\ref{fig:PSAspectrum}, where the high spectrum is all events from a \nuc{232}{Th} source, and the low spectrum is for events that pass the PSA cut. The strong peak remaining is the double-escape peak from the 2615-keV \nuc{208}{Tl} $\gamma$ ray, which is a proxy for single-site \BBz-decay events. The algorithm retains at least 95\% of these events, while rejecting up to 99\% of the single-escape, multi-site events. One should compare this to the A/E results of Ref.~\cite{dusa09} where the double escape peak events are accepted at 89\% and the single-escape peaks are rejected at 93\%.

More recently we have also developed a model for slow, partial-energy events from interactions within the lithium contact layer on the outer edge of \ppc\ detectors. This layer is typically 1 mm in thickness, but is not entirely inactive material. Hence, events within that layer can produce signals with long rise-times and partial energy collection. We now understand these signals as the result of competition between the diffusion of holes out of the Li layer with the recombination of electrons and holes at Li precipitates~\cite{Aguayo2013,Finnerty2013}. Since these slow events can potentially generate much of the background at very low energies, a detailed and comprehensive understanding of this process is crucial for dark matter, axion and other low energy-dominated physics sensitivity.

\section{The \MJDEMbf\ Construction and Facility}
\subsection{Enrichment, Ge reduction and refinement, and detector production}
The \DEM\ baseline plan calls for 30 kg of enriched Ge detectors. The Collaboration acquired 42.5~kg of $^{76}$Ge in the form of 60.5 kg of $^{76}$GeO$_2$, which was produced by the Joint Stock Company Production Association Electrochemical Plant (ECP) in Russia.  The order was delivered to Oak Ridge, TN, United States in two shipments. The first 20~kg was delivered in September 2011, while the rest was delivered in October 2012.  A special steel container was constructed to minimize the exposure of the enriched \nuc{76}{Ge} to cosmic rays during transport.  The calculated cosmic ray production of \nuc{68}{Ge} and \nuc{60}{Co} is reduced by a factor 10 and 15, respectively, for samples transported within this container. Shielded storage for the enriched material being processed in Oak Ridge is provided by a cave located about 8~km from the processing and detector manufacturing facilities. The cave has an overburden of 40~m of rock, which is more than adequate for shielding the enriched material from the hadronic component of cosmic rays.

Electrochemical Systems, Inc. (ESI), in Oak Ridge, TN provided the first stage of material preparation.   Before processing any enriched material, pilot tests with natural GeO$_2$ were performed to qualify the procedures.  The delivered $^{76}$GeO$_2$ from ECP underwent a high temperature reduction in a hydrogen atmosphere.  When the resistivity of the reduced material was greater than 3 $\Omega$-cm, the material was then purified by zone refinement to reach a resistivity of $>$47 $\Omega$-cm.  The processing by ESI provided a yield of 98\% of enriched material suitable for further refinement and detector manufacturing. In addition to the process qualification of the refinement at ESI, AMETEK/ORTEC produced two \ppc\ detectors fabricated from natural Ge that had been reduced and purified by ESI. 

The refined enriched material from ESI was further purified by zone refining at AMETEK/ORTEC before being used as charge in a Czochralski crystal puller.  Detector blanks were cut from the pulled crystals ($\sim$70-mm diameter), followed by the standard detector manufacturing steps of lithiation, implantation of the p$^+$ contact, and passivation.  At each step, the mass of the enriched materials being handled was recorded for inventory control.  In the production of the enriched \ppc\ detectors, slurries from detector cutting and shaping processes, as well as small samples cut from the pulled crystal for evaluation, were saved and reprocessed by ESI for reuse in detector production.  Acids that were used in the detector manufacturing process were not saved.  At all stages of the detector fabrication process, any enriched materials that were not being worked on were transported back to the cave for storage.

Once an enriched detector is manufactured, it is mounted in a PopTop\texttrademark~ capsule and tested for performance.  The PopTop\texttrademark~ detector capsule, which can be detached from the cold finger, offers superior portability and ease of disassembly --- qualities that are essential for transferring the detector from the capsule to the low-background string mounts used in the \DEM.  In order to mitigate contamination of the mounted detector in a PopTop\texttrademark~ capsule, the parts that are in direct contact with the detector are made from only radiopure materials.  For example, indium contacts are replaced by gold or clean tin contacts, and synthetic charcoal with low radon emanation rate is used to maintain the vacuum in the capsule.  

The production of enriched detectors began in November 2012.  As of April 2013, ten enriched detectors with a total mass of approximately 9.5~kg have been fabricated with eight delivered to SURF.  The detectors were transported by ground to SURF, and a portable muon counter~\cite{Aguayo2013b} was used to log the cosmic-ray exposure during the trip.

\subsection{Detector Array Configuration}
The detector array for \DEM\ is designed with many goals in mind. The functional requirements are:

\begin{itemize}
\item Only the most radiopure materials are used to construct the detector holder.
All the detector and string components are made out of two possible materials: underground electroformed Cu (UGEFCu) from the Temporary Clean Room (TCR) (Sec.~\ref{Sec:electroforming}) or NXT-85 (a teflon that is specially manufactured in a cleanroom environment, Sec.~\ref{Sec:Background}). The UGEFCu has a maximum thickness of 1.27 cm and the NXT-85 parts are fabricated from 15.875-cm long rods that are 3.175 cm in diameter. The NXT-85 mass is minimized and used only where electrical insulation is required.
\item UGEFCu and NXT-85 parts are processed in the cleanroom machine shop (Sec.~\ref{Sec:UGFacilities}), so designs must be compatible with the machine tools purchased for this shop. Use of wire Electric Discharge Machining (EDM) is preferred as a clean material removal technique.
\item A 5-mm vacuum gap or 1 mm of NXT-85 is required to isolate high voltage from neutral components.
\item The detectors have variable dimensions in order to maximize the yield of enriched germanium.  Unique parts are minimized in order to allow a wide range of detector sizes to be packaged, while still providing a high packing factor of germanium in each cryostat.
\item Threaded connections are difficult to produce and keep clean. We minimized the number of threaded connections, and where that was not practical, special methods are employed to ensure quality, cleanliness, and repeatability of threaded connections.
\end{itemize}

\begin{figure}[!htbp]
\begin{center}
\includegraphics[width=7cm]{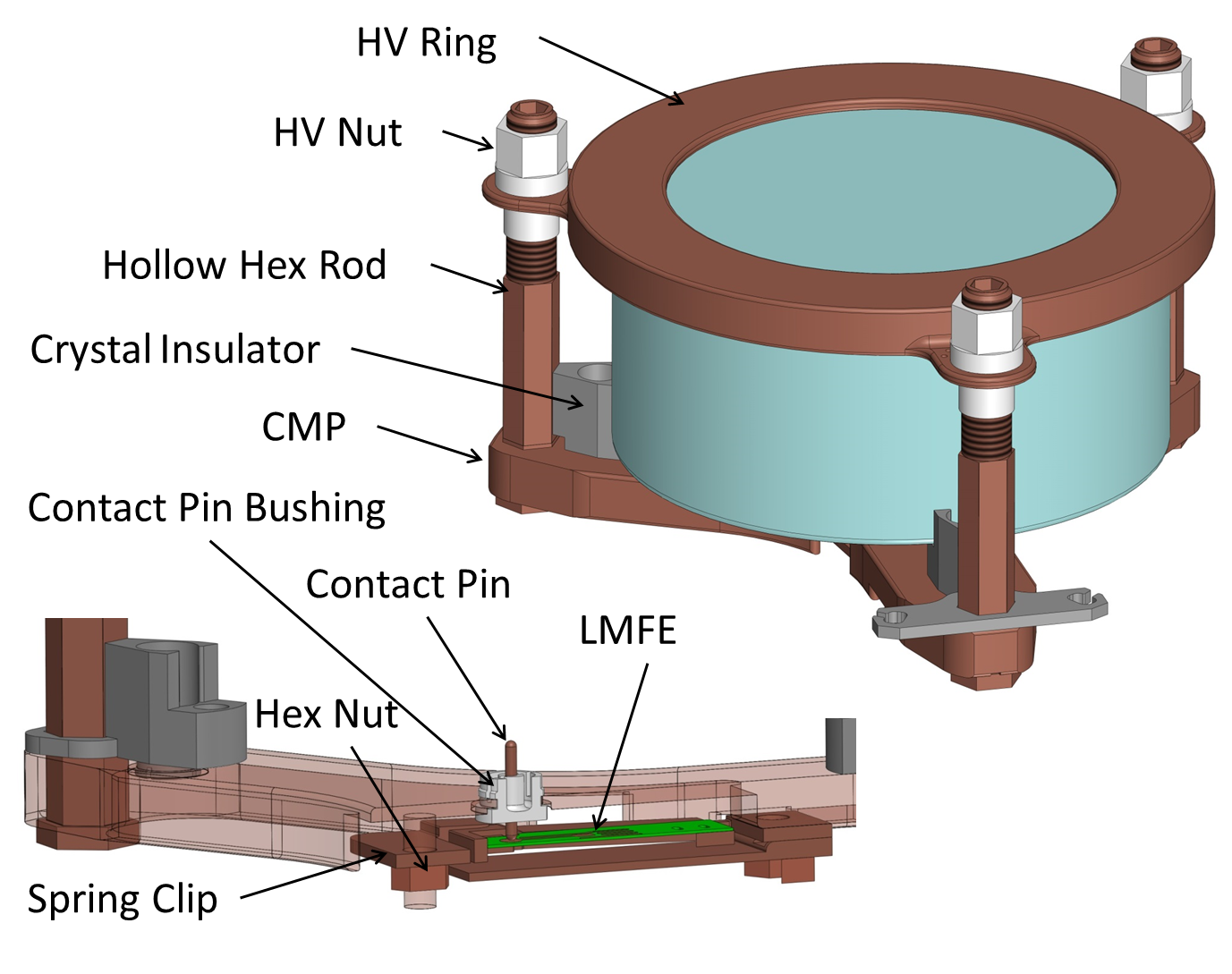}
\caption{A rendering of the detector unit design. See text for details.}
\label{fig:DetectorUnit}
\end{center}
\end{figure}

Each detector is housed in a frame referred to as a detector unit (Fig.~\ref{fig:DetectorUnit}). The crystal mounting plate (CMP) is the foundation, while 3 hollow hex rods and High Voltage (HV) nuts provide connection to the  HV ring, which clamps the detector in place. The crystal insulators provide electrical isolation between the HV surface of the detector and the neutral CMP. The crystal insulators are also sized to compensate for the different coefficient of thermal expansion of copper and germanium, providing equivalent clamping force when warm and cold.

The crystal insulators snap into place on the CMP, as does the contact pin bushing. The contact pin and Low Mass Front End (LMFE) board are held in place by the spring clip, which provides contact pressure between the pin and detector. The spring clip is held in place and tensioned by parylene coated \#4-40 Cu nuts on threaded milled studs in the CMP. Parylene coating of these nuts provides a higher strength connection than NXT-85 nuts, while providing thread lubrication to prevent copper galling.

Up to five detector units are stacked into a string. The string is clamped together with tie rods. The string adapter plate connects the string to the coldplate. The tie rod bottom nuts and adapter plate nuts, also parylene coated, provide a strong clamping force for thermal contact.

\begin{figure}[!htbp]
\begin{center}
\includegraphics[width=7cm]{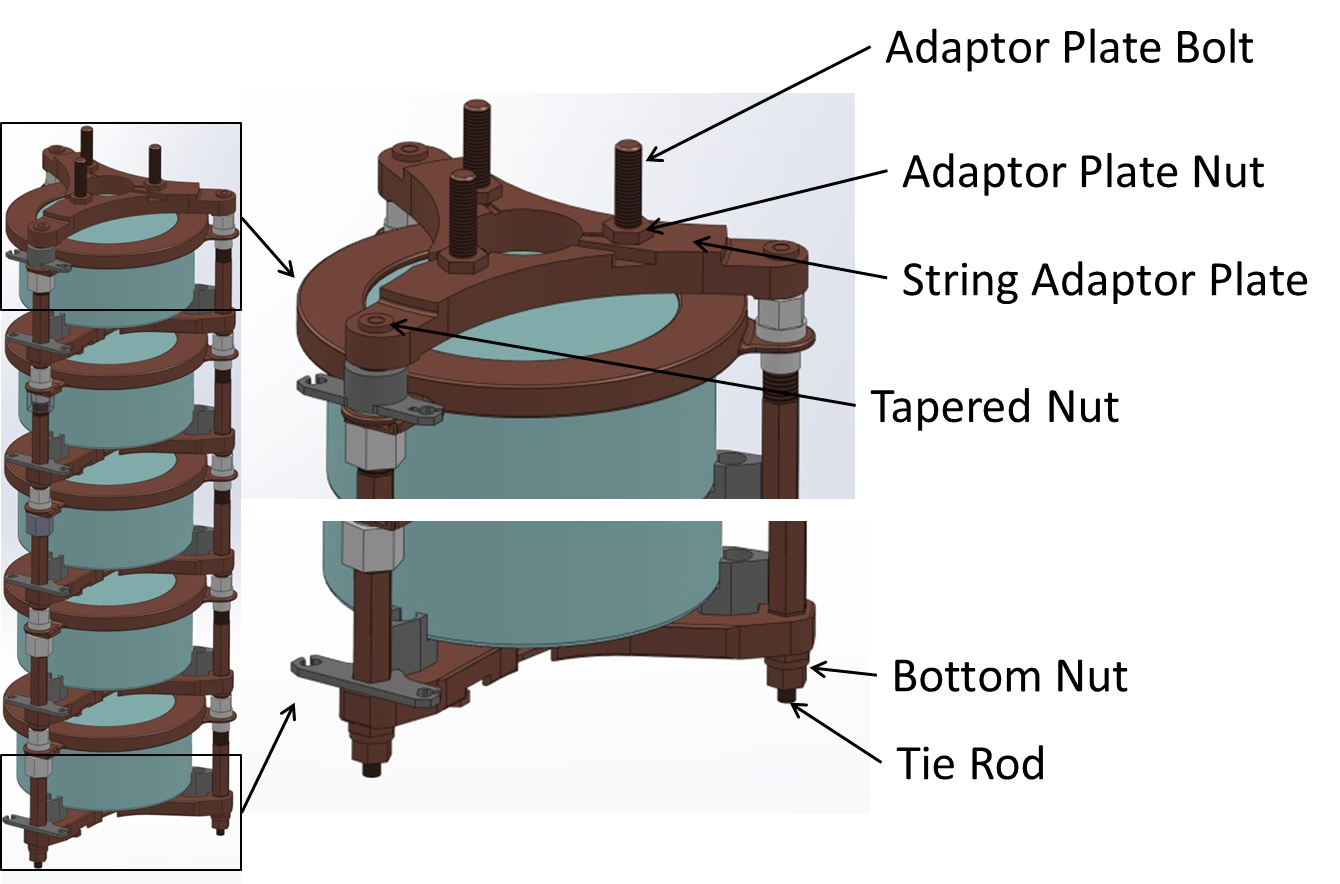}
\caption{A rendering of the string design. See text for details.}
\label{fig:DetectorString}
\end{center}
\end{figure}

The LMFE board is built on a 0.025-cm thick fused silica substrate, with all of the cold electronics mounted on the board. The LMFE assembly is then inserted into spring fingers on the spring clip. This assembly is then mounted to the CMP. After the contact pin and detector are mounted, the spring clip tensioning nut is set to apply the appropriate contact pressure.

The detector units are designed to accommodate a wide range of detector sizes from multiple vendors. Detectors have the shape of a right cylinder, and can have a diameter of 50-77 mm, and a height up to 65 mm. Variations in diameter and corner sharpness can be accounted for by producing custom crystal insulators. Variations in the details of the pin contact with respect to the overall shape are accounted for by having 2 different pin lengths. Variations of less than 0.5 mm in this geometry can be accounted for by adjusting the spring clip tension starting point.

The special methods for threaded connections include the use of dedicated tools, as with all parts machined underground, to avoid cross contamination. The \#4-40 studs on the CMP are thread-milled from bulk material, so instead of having independent screws, there are in-place studs that protrude from the interior surface. Interior threads are made using roll-form taps, which produce a consistent thread quality and no burr. All interior threads are of the smallest depth necessary for strength, and there are no blind tapped holes. This makes cleaning and drying parts easier. All threaded parts are verified by hand before release for final cleaning.

\subsection{The Electroformed Copper Cryostats, the Thermosyphon, and the Vacuum System}

\subsubsection{The \MJDEMit\ Module}

	The \MJ\ \DEM\ is a modular instrument as the detector strings are deployed in two copper cryostats, each outfitted with its own vacuum and cryogenic systems for independent operation.  This modular scheme allows for phased deployment of detectors as they become available, and suggests a scheme for development of a tonne-scale \nuc{76}{Ge} experiment; a larger experiment can be constructed by deploying several similar cryogenic modules.  The cryostats designed for the \DEM\ are each capable of housing seven of the previously described detector strings, for a total capacity of $\sim$20 kg of HPGe detectors apiece. The first cryostat, Cryostat 1, will contain detectors produced from both natural and enriched germanium. The second cryostat, Cryostat 2, will only contain detectors produced from enriched germanium. The cryostats are constructed from copper: a design decision motived by the ability to produce ultra-pure copper through chemical electroforming. Cryostats 1 and 2 are fabricated from this ultra-pure copper, while an initial prototype cryostat is fabricated from commercially-sourced copper. The Prototype Cryostat will only contain two strings of detectors produced from natural germanium. It serves as a testbed for mechanical designs, fabrication methods, and assembly procedures that will be used for the construction of the electroformed-copper Cryostats 1 \& 2.
	
\subsubsection{Electroforming}
\label{Sec:electroforming}

The primary requirement for the copper used in the \MJ\ \DEM\ is that it is sufficiently purified.  This includes removal of naturally occurring radioactivity from U and Th, as well as the elimination and prevention of reformation of cosmogenic radioisotopes species.  Due to its large total mass, the radiopurity goal for the copper that is used in the inner shield and detector components is very stringent.  To attain the background goal of \threecpRty, the required purity levels are $<0.3~\mu$Bq \nuc{238}{U}/kg Cu (or $2.4\times10^{-14}$ g \nuc{238}{U}/g Cu) and $<0.3~\mu$Bq \nuc{232}{Th}/kg Cu (or $7.5\times10^{-14}$ g \nuc{232}{Th}/g Cu). Electroforming copper in a carefully controlled manner within a clean environment allows one to produce copper with the required radiopurity~\cite{Hoppe2009}. 

A secondary requirement for the electroformed copper relates to its physical properties.  The mechanical properties of electroformed copper can vary drastically depending on the conditions under which it was formed. Conditions that favor high purity can form large crystalline structures with poor mechanical strength. Small polycrystalline formations can exhibit adequate tensile strength but lower purities. These conditions, which are seemingly at odds with one another, require that a careful balance of operational parameters be obtained in the electroforming production process.  

Design considerations for load bearing components were carried out using conservative estimates for material properties such as yield strength.  The design yield stress value used for electroformed copper was estimated to be 48 MPa.  Mechanical testing and evaluation is necessary to prove the plated material's ability to withstand the loading conditions without failure. Mechanical evaluation has shown the yield strength to be 83.2 MPa~\cite{Overman2012} on average with a significant degree of strain hardening observed. The UGEFCu has, therefore, shown compliance with the design criteria.

The electroformed material for the \DEM\ has been fabricated mostly from cylinders that are up to 35.6 cm in diameter (the inside diameter of the cryostats). The thermosyphon was formed on a mandrel 1.90 cm in diameter. The copper produced can range from a few tens of microns to very thick plates near 1.4 cm. Time constraints are the primary limitation when producing very thick electroforms.  The current plating rate for the \DEM's copper is typically 38 to 64 $\mu$m per day, depending on a variety of parameters.  While this rate can be increased, it is at the expense of purity and mechanical properties of the electrodeposited material.  From the plating rate indicated, a 1.4-cm thick electroform takes approximately 8-12 months to complete.

\subsubsection{Cryostat Design}

\begin{figure*}[!htbp]
\begin{center}
\includegraphics[width=12cm]{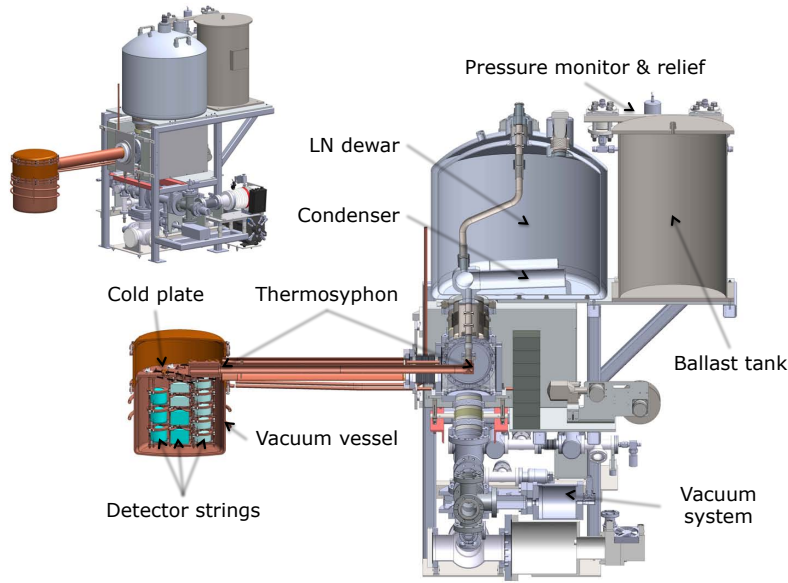}
\caption{The \MJ\ \DEM\ Module.  Detector Strings are housed within ultra-low background cryostats, each of which are supplied with their own vacuum and cryogenic systems.  See text for details of vacuum and cryogenic system function.} 
\label{fig:Cryostat}
\end{center}
\end{figure*}

	The cryostat is a copper vacuum enclosure that an electron-beam-welded vessel assembly along with removable top and bottom lids (see Fig.~\ref{fig:Cryostat}).  The vacuum seals between these components are a custom \MJ\ design that uses thin (51 $\mu$m-thick) parylene gaskets sandwiched between tapered surfaces machined into the copper components.  Copper rail sectors and clamp bolts are used to maintain parallelism during assembly and pump down.  Vacuum forces are sufficient to maintain the seal, so bolt strength is not a factor in effective sealing.  
	
	Detector strings are mounted to a copper cold plate, which rests on Vespel\textregistered ~pins that provide support and alignment, while maintaining a thermal break from the room temperature vacuum vessel.  An infrared (IR) shield is mounted to the underside of the cold plate to reduce detector leakage current generated by IR radiation. 
	The cryostat is supported at its crossarm by a copper frame (not shown in Fig.~\ref{fig:Cryostat}) inside the lead stack.  A transition is made to stainless steel conflat vacuum hardware at the far end of the crossarm tube via a copper/stainless explosion-bonded transition flange.  All stainless steel vacuum hardware is located outside of the \DEM's passive shielding.

\subsubsection{Vacuum System}

	Each cryostat is mounted to its own vacuum system, constructed from all-metal ultra-high vacuum components.  A 200-lpm, oil-free diaphragm pump provides rough vacuum, a 300-lps turbo-molecular pump is used for initial pump-down to UHV pressures, and a 1500-lps cryopump is used for steady-state operation.  A non-evaporable getter (NEG) pump is used to remove built-up noncondensable gasses.  A residual gas analyzer provides mass spectrometry analysis of the vacuum.  All of the active components, including valves, are remotely-operable, and an application has been developed to monitor and control the system, allowing full remote operation. (See Sec.~\ref{Sec:DAQ}.)  Pressures are continuously uploaded to a slow-controls database for history viewing.

\subsubsection{Cryogenics}

	The detector strings are cooled via the cold plate by means of a thermosyphon~\cite{Aguayo2013a}. The thermosyphon is a closed tube within the crossarm that joins the cold plate to a condenser volume residing outside the shield and inside a dewar containing liquid nitrogen. The thermosyphon contains nitrogen that is liquefied at the condenser and transported by gravity down the length of the crossarm to the cold plate, where, as it evaporates, it cools the cold plate.  The evaporated nitrogen travels back to the condenser where it is re-liquefied. In this cycle, heat is transported from the cold plate to liquid nitrogen in the dewar. The liquid nitrogen in the dewar then evaporates and is replenished from an external supply. The dual phase nitrogen within the thermosyphon has a large effective thermal conductivity providing the required cooling power for the \DEM. The operating temperature can be tuned by adjusting the amount of thermosyphon nitrogen.  By producing only a thin layer of condensed nitrogen at the cold plate, microphonics from evaporation are minimized. 
	
	The thermosyphon system consists of a thermosyphon tube, custom liquid nitrogen dewar, gaseous N$_2$ plumbing for loading nitrogen into the tube, and an external ballast tank.  The thermosyphon tube is constructed from the same grade of UGEFCu as that from which the cryostat is fabricated.  The liquid nitrogen dewar is a custom-fabricated device which includes the condenser volume with an in-vacuum connection to the thermosyphon tube and an in-air connection to the N$_2$ supply.  The external ballast tank is primarily a safety feature, allowing for evaporation of condensed thermosyphon-tube nitrogen in the case of a loss of liquid nitrogen supply, without creating a hazardous over-pressure condition.  Additionally, since the ballast tank is isolated from the thermosyphon tube, nitrogen can be stored for several \nuc{222}{Rn} half-lives (3.8 days) before it is loaded into the thermosyphon tube. In this way we can ensure that the nitrogen circulated in the thermosyphon tube is radon-free.

\subsection{Detector Acceptance, Characterization and Calibration}
There are a number of experimental characteristics that need to be monitored during the course of the experiment including:

\begin{itemize}
\item Energy scale and linearity
\item Absolute efficiency for double-beta decay within each detector
\item Energy resolution and  peak shape
\item Background and signal tagging efficiencies
\item Pulse waveforms response
\end{itemize}

We have developed a three-phase plan to ensure the required data are acquired. The phases include acceptance, characterization, and calibration. The acceptance testing is the initial phase of evaluating the detector performance.  This type of testing is done upon receipt of the detectors from their manufacturers, and is performed in the transport cryostats with the original manufacturer-supplied preamplifiers.  The tests conducted during this phase are relatively cursory, and meant solely to establish that the received detectors meet some minimum qualifications.  The basic tests performed are: energy scale and resolution, relative efficiency, leakage current or capacitance at depletion, initial estimate of the dead layer, and detector mass and dimensions. 
	
Characterization measurements are conducted to fully determine the operating behavior of a detector. This includes: energy scale, resolution, capacitance measurements, single site and multiple site event separation performance, dead-layer measurements, and crystal-axis measurements. Characterization measurements are done with detectors in the final string configuration both within a test cryostat and within the \DEM\ cryostat prior to operation, {\it{i.e.}} prior to commissioning.
	
The calibration measurements are designed to monitor the stability of the system during run-time operation.  Initially, hour-long calibrations with a \nuc{228}{Th} source will be conducted on a weekly basis.  These runs will measure the stability of the energy scale and resolution, efficiency, and pulse-shape analysis (PSA) efficiency.  Once the stability of the detectors has been established, the time period between the source calibration can be extended to twice monthly or even monthly. 

In the detector and string characterization stage, measurements are being performed with button sources (\nuc{133}{Ba}, \nuc{60}{Co}, \nuc{241}{Am}) in a clean room environment.  Once the detector strings are loaded into the cryostat, access will be limited. Each monolith will have a low-background source pathway of PTFE tubing that spirals around the outside of the cryostat.  A line source of \nuc{228}{Th} will be remotely fed into the pathway, enabling the calibration of the entire cryostat with a single source either during final testing or after the monolith is placed in the shield.  Simulations have shown that in an hour-long run we can accumulate the necessary statistics to monitor the efficiency, PSA performance, energy scale, and resolution, while simultaneously keeping the count-rate below the signal pile-up threshold of $\sim$100 Hz.  The source will be parked in an external garage separated from the shield-penetrating section of the pathway by an automated valve system. During calibration runs the valve will be open, and the entire pathway will be purged with liquid nitrogen boil-off.  During production runs, the shield-penetrating section of the pathway will be sealed off from the garage. The source itself will be encased in two plastic tubes to prevent leaving residual radioactivity within the pathway. 

\subsection{Electronics and Data Acquisition}
\label{Sec:DAQ}
\subsubsection{The Detector Readout Electronics}

\begin{figure}[!htbp]
\begin{center}
\includegraphics{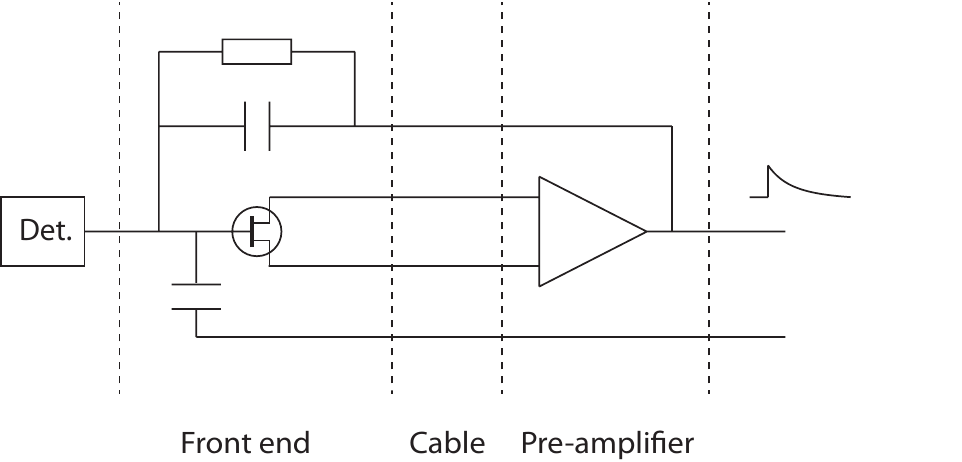}
\caption{A high level illustration of the signal readout scheme for each detector channel. See text for details on this readout design and the typical values for the components.} 
\label{fig:SigProcess}
\end{center}
\end{figure}

Each of the two cryostat modules in the \DEM\ contains seven strings, with each string holding up to five detectors.   Figure~\ref{fig:SigProcess} illustrates the basic design of the low-noise, low-radioactivity signal-readout electronics.  It consists of the LMFE~\cite{Barton2011}, a circuit containing the input FET and feedback components, which is located close to the detector in order to minimize stray input capacitance. It also includes the preamplifier, which lies outside the cryostat and is connected to the LMFE by a long cable. The main challenges to the practical realization of this design are sourcing components for the front end that are low in both noise and radioactivity, and dealing with the long cable in the feedback loop.

The LMFE is a resistive feedback circuit. This architecture has various benefits over the common pulse reset alternative; namely that it is simpler and avoids interference from reset pulses: a problem for multiple-detector systems.   The LMFE is 20.5$\times$7 mm$^2$ in size and weighs approximately 80 mg. The substrate is fused silica, which was chosen for its high radiopurity, low dielectric losses, and low thermal conductivity. Its low thermal conductivity means that a significant temperature gradient can be maintained across the board.  By choosing an appropriate geometry and thermal conductivity for the substrate, the FET can be maintained at its optimal operating temperature for noise performance by self-heating which is adjustable by controlling drain to source voltage.  The chosen FET is  a bare Moxtek MX-11 JFET low-noise die with high transconductance and very low input capacitance. It is attached to the silica board with silver epoxy via its gate substrate, while the source and drain pads are connected to traces with Al-Si wire bonds.
The feedback resistor is formed by sputtering a layer of amorphous Ge (a-Ge) and has a resistance of $\sim$10-100~G$\Omega$ at cryogenic temperatures.  The feedback capacitance of $\sim$0.2~pF is formed by the stray capacitance between traces on the front end.  

Extraordinarily low noise levels can be achieved using the front end design previously described.  The equivalent noise charge achieved without a detector was $55$~eV FWHM, and with a detector, a small \ppc, it was $85$~eV. The noise levels are not representative of what will be achieved in the \DEM\ because of the greater detector capacitance, but they indicate that the noise resulting from the materials and components of the front end board will not be the limiting factor.  To minimize the material budget and thermal dissipation in the cryostat, a 0.4-mm diameter, $\sim$2-m long miniature 50-$\Omega$ coaxial cable is used to drive signals from the LMFE to the first stage of the preamplifier, based on a classic folded-cascode transistor design.  The preamp is capable of rise times below 10 ns with judicious choice of front end components and a short cable connecting it to the front end. With the long cable used in the \DEM\ configuration, this rise time increases to $\sim$40-70~ns, depending on the length of the cable.

\begin{figure*}[!htbp]
\begin{center}
\includegraphics[width=12cm, angle=-90]{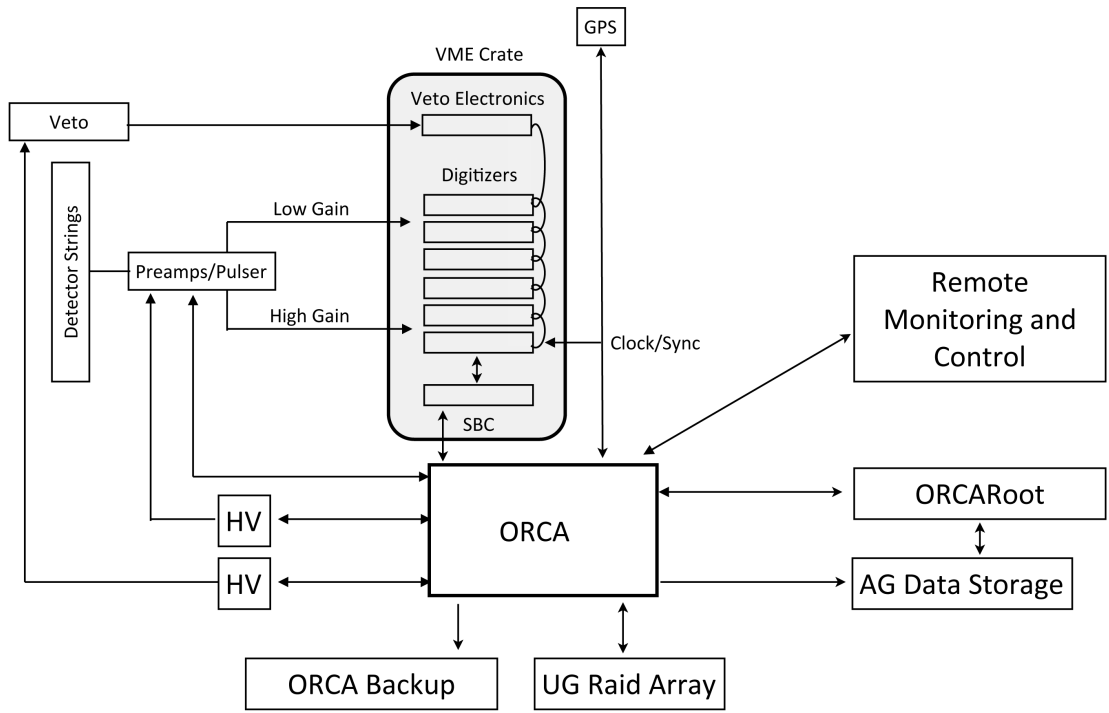}
\caption{A high level schematic of the DAQ system. Each detector uses a low- and a high-gain digitizer channel for a total of 70 channels/crate. Each VME crate has one Single Board Computer (SBC) that reads out all electronics in its crate, sending the resulting data to ORCA. Data is buffered locally underground and then transferred to the surface. There will be one VME crate per module. The Veto electronics will reside only in one of the two crates.}
\label{fig:DAQFlow}
\end{center}
\end{figure*}

The preamplifiers are organized by string position on motherboards.  On the motherboard, each detector has a low-gain and a high-gain signal output for digitization, resulting in a maximum of 70 channels per cryostat. The digitization electronics for each cryostat are in separate VME crates. Each crate has a single board computer to read out the digitizers in that crate and the entire system is controlled by a central DAQ computer. A conceptual schematic of the data-acquisition electronics is given in Fig.~\ref{fig:DAQFlow}.

The collaboration is using GRETINA digitizer boards~\cite{Zimmermann2012} developed as part of the GRETINA experiment$^*$\footnotetext[1]{We gratefully thank the GRETINA collaboration for digitizer board loans.}.  The GRETINA cards are a combination of a digitizer and digital signal processor, which accept 10 inputs directly from the detector  pre-amplifiers and digitize at a nominal frequency of 100 MHz with 14 bit ADC precision.  A Field-Programmable Gate Array (FPGA) performs digital leading edge and/or constant fraction discrimination, trapezoidal shaping, and pole-zero correction.  Some special firmware extensions were specifically developed for the \DEM. These modifications allow for independent configurations of the 10 input channels that allow the raw traces from each detector to be decimated in such a way as to simultaneously capture a fully sampled rising edge with pre-summed regions before and after the edge for studying slow pulses. The capability of measuring the input trigger rate (triggers/second) for each individual input channel is another feature added to the card specifically for the \DEM. Data records are fixed at 2020 samples/event and the card allows for these samples to be the sum of various ADC output values. This summing capability allows the card to achieve selectable time windows (20, 40, 80, 160, and 200 $\mu$s) for the acquired data while maintaining a constant event length. In the case of the \DEM, different crystal geometries can be plugged into the same card, and therefore it is mandatory to be able to accommodate the different time constants of the different crystals.
  
A controller card communicates between the motherboard and the digitization electronics. The controller card contains 16 pulse generator outputs with individually controlled amplitudes, 16 DAC levels outputs for setting the drain-to-source voltage of the FETs, and 16 ADC inputs to monitor the first stage outputs of the preamplifiers. The pulser outputs of the controller card are used for electronics calibration, monitoring the gain stability of independent channels, and monitoring the trigger efficiency. The controller cards are implemented in the overall DAQ system and slow control processor.

\subsubsection{Other Data Acquisition System Electronic Systems}

The passive lead shield is surrounded by scintillator panels that are used for an anti-coincidence (veto) shield. The veto (see Sec.~\ref{Sec:Shield})  has separate electronics in one of the VME crates with some additional electronics in a separate NIM BIN. Data from the veto is read out by the main DAQ software and integrated into the detector data stream. All veto events are time stamped using a scaler that counts the common clock pulses. 

To ensure accurate time-stamping of the signals, a common clock from a Global Positioning System (GPS) module is distributed among the digitizer cards and the veto system. A  common reset is used as a system synchronization pulse to simultaneously reset counters on all digitization and veto boards, thus providing an absolute time reference for each event.

Separate computer-controlled HV bias supply systems are used for the detector array and the photomultipliers of the veto shield. Each detector and phototube is powered by an independent HV channel, allowing for optimization of the HV setting for each detector, and enabling any detector to be taken offline without affecting the rest of the array.

Data are transferred from the DAQ computer to an underground 32 TB RAID system, and from there to the above-ground analysis computer system. The underground RAID storage is used as a buffer to prevent loss of data in the case of an underground-to-surface network failure.

\subsubsection{The Data Acquisition System Software}

The DAQ software system used by the \DEM\  is the Object-oriented Real-time Control and Acquisition (ORCA) \cite{How04} application. ORCA is a general purpose, highly modular, object-oriented, acquisition and control system that can be configured at run-time to represent different hardware configurations and data read-out schemes by dragging items from a catalog of objects into a configuration window. Since each object is comprised of its own fully encapsulated data structures as well as support and diagnostic code, ORCA can easily support specific experiments, such as the \DEM.

\subsection{The Shielding Configuration}
\label{Sec:Shield}
\begin{figure}[ht]
\begin{center}
\includegraphics[width=7cm]{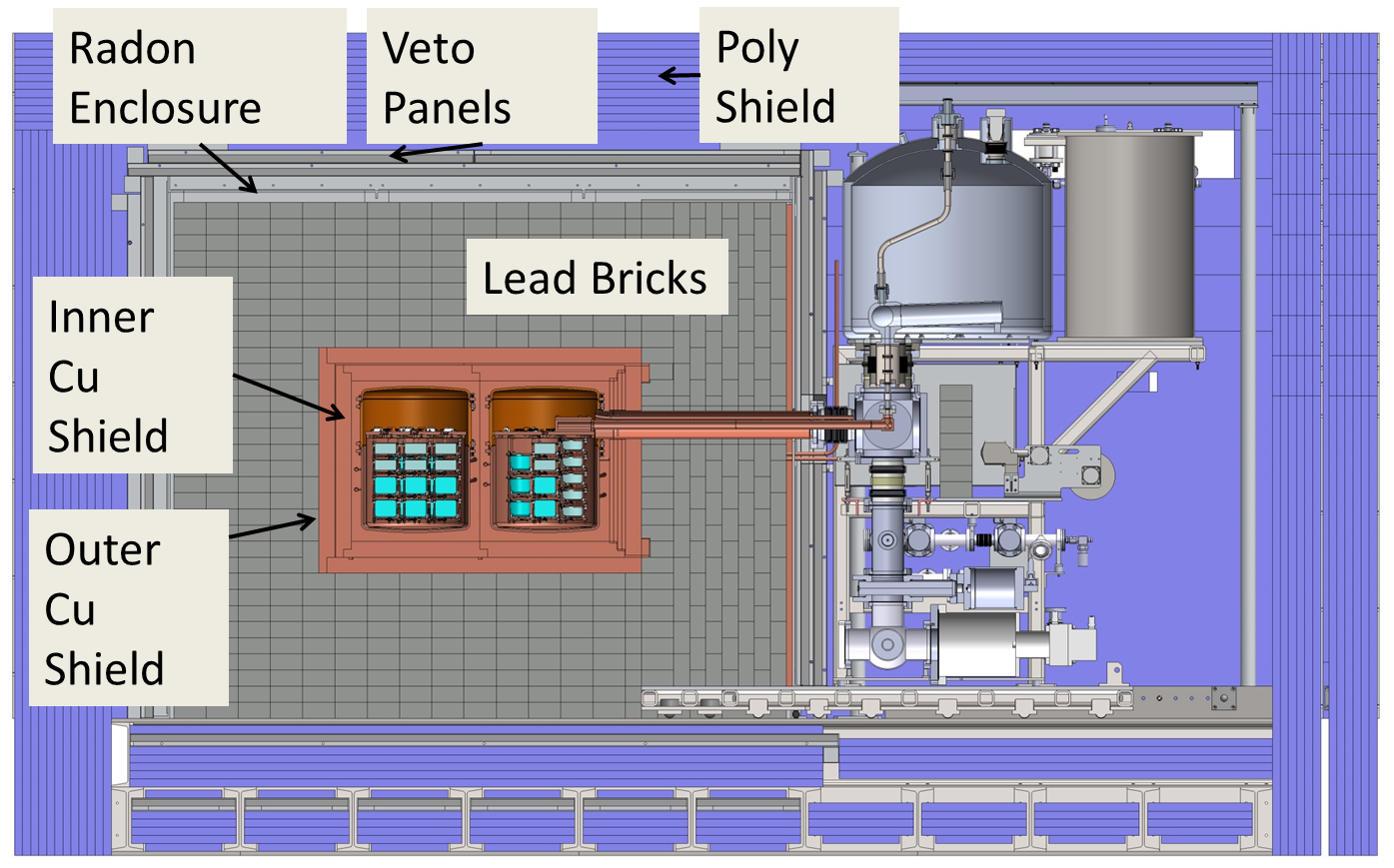}
\caption{The shield system in cross section, shown with both cryostats installed.}
\label{fig:ShieldOverview}
\end{center}
\end{figure}

 The passive shield consists of graded shield materials starting outside of the cryostat and extending out to the overfloor table, or base plate, upon which the experiment is assembled. This system also includes the integral calibration source track and drive system, the transport mechanisms for installing and retracting the cryostat from the bulk of the shield for access, and a radon scrubbing and nitrogen delivery system for providing purge gas to the internal portions of the shield. A high-level summary of shield components is given in Table~\ref{tab:ShieldComponents}, and the complete shield assembly is shown in Fig.~\ref{fig:ShieldOverview}.

 \begin{table*}[htdp]
\caption{A summary of the shield components.}
\begin{center}
\begin{tabular}{|l|l|l|}
\hline
Shield Component 		& Material		 				& Thickness		\\
\hline\hline
Inner copper shield		& 4 layers 1.25-cm thick UGEFCu	& 5 cm			\\
Outer copper shield		& OFHC copper, commercial		& 5 cm			\\
Lead shield				& $5.1 \times 10.24 \times 20.3$ cm$^3$ bricks	& 45 cm \\
Radon exclusion box		& Al sheets 						& $0.32 - 0.635$ cm		\\
Veto panels				& Scintillating acrylic			& 2 layers, 2.54 cm each	\\
Poly shield				& HDPE							& 30 cm, inner layer borated \\
\hline
\end{tabular}
\end{center}
\label{tab:ShieldComponents}
\end{table*}%

Gamma rays from the inner region of the shielding contribute to the background. Therefore, materials with extremely low radioactivity must be used in this region. Copper can be purchased very pure and made ultra-pure via electroforming (Sec.~\ref{Sec:electroforming}) to meet the specifications required by the background budget. The innermost layer of the shield will be 5 cm thick and built of UGEFCu sheet. The practical upper limit on the thickness for electroforming in our baths is about 1.40 cm, so this shield is constructed of four layers of 1.25-cm-thick plates.
 
 The outer copper shield is made of 5-cm-thick OFHC copper plates, and machined and bolted together to provide the mechanical reference points for the experiment. This shield ``box'' is supported off the overfloor table on four OFHC copper legs to allow for precision alignment and reference independent of the lead shield bricks. The lead shield consists of a 45-cm-thick layer of 5.1 $\times$ 10.2  $\times$ 20.3 cm$^3$ bricks, to be stacked in place and machined as needed. 
 
 The copper and lead shield will be contained inside a semi-sealed aluminum box. This box will permit the controlled purging of the gas within the inner cavity that contains the detector modules. The box is constructed of a welded and bolted aluminum structural frame, with 0.3175-cm thick bolted Al panels. All seams and openings are sealed, with the exception of the seals around the monolith, which are gasketed and bolted for easy removal and replacement.
 
 Two layers of veto panels surround the radon exclusion box. Each panel consists of a 2.54-cm-thick scintillating acrylic sheet. This sheet has small longitudinal grooves machined for wavelength-shifting fibers, and is wrapped in a custom-made reflecting layer to compensate for light attenuation along the fibers. Light from the fibers is read out by a single 1.27-cm photo-multiplier tube. The scintillator assembly is enclosed in a light-tight aluminum box. There are a total of 32 veto panels surrounding the radon exclusion box, including a number that reside within openings of the overfloor. Sheets of 2.54-cm-thick high density polyethylene (HDPE) panels, stacked up to 30 cm of total thickness, make up the poly shield structure. The inner 2 layers consist of borated HDPE.

\subsection{Underground Facilities}
\label{Sec:UGFacilities}

The Homestake Mine is home to the Sanford Underground Research Facility (SURF) that has been developed by the state of South Dakota as a site for experiments requiring underground laboratory space. The state of South Dakota, along with private donor T. Denny Sanford, have committed funds that have allowed refurbishment and access to underground space. Operation of SURF is funded by the US Department of Energy and the State of South Dakota.   

\begin{figure*}[!htbp]
\begin{center}
\includegraphics[width=15cm]{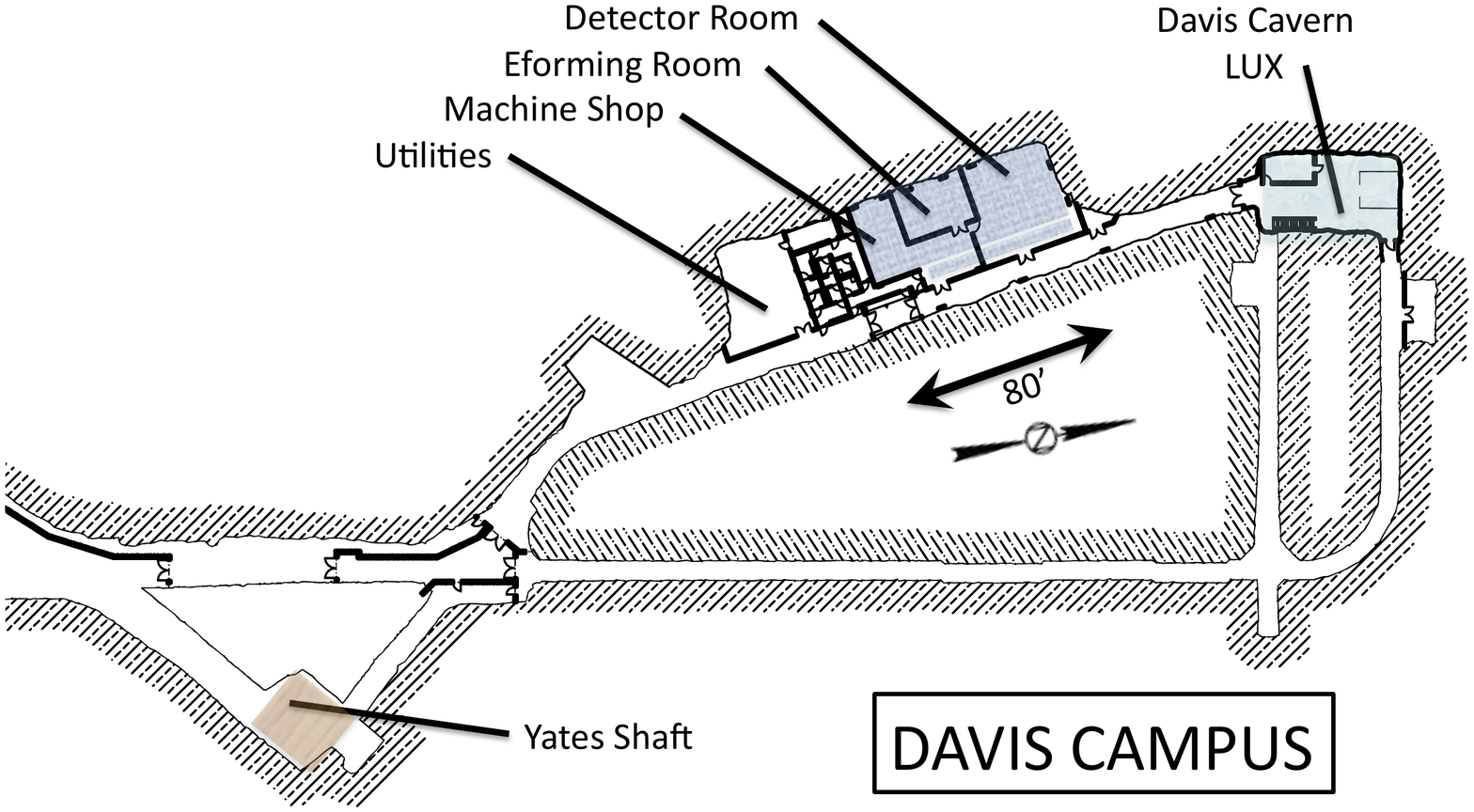}
\caption{The layout of the Davis Campus at SURF showing the three \MJ\ labs and the location of the LUX experiment at the Davis Cavern. The corridor that joins the two projects is a clean space.}
\label{fig:DavisCampus}
\end{center}
\end{figure*}

\begin{figure*}[!htbp]
\begin{center}
\includegraphics[width=16cm]{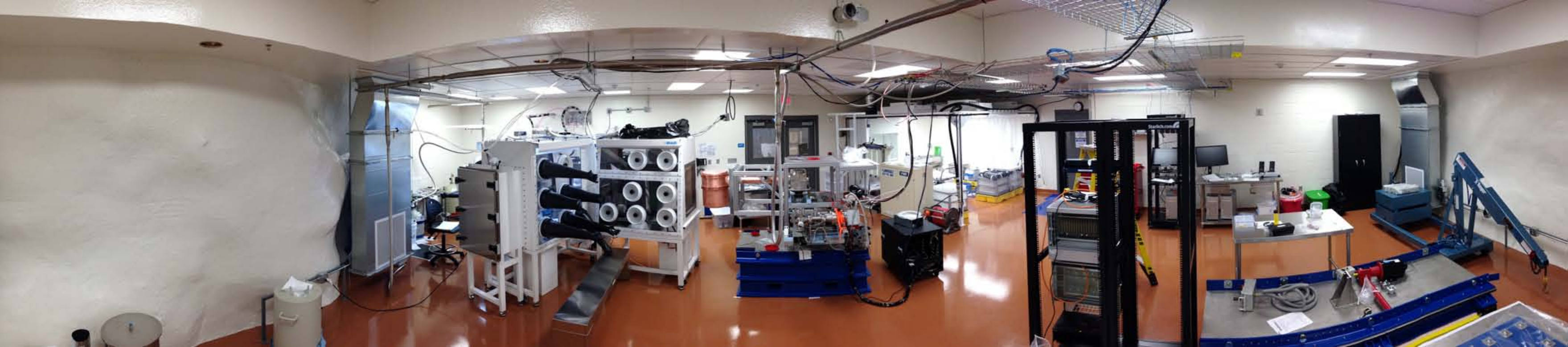}
\caption{A view of the detector laboratory at the Davis Campus at SURF.  In this view, the glove box is to the left, and the prototype module is center. In the rear, a soft-wall cleanroom is shown that serves the purpose of an area of increased cleanliness.}
\label{fig:LabLayout}
\end{center}
\end{figure*}

The \MJ\ laboratory space at SURF consists of three cleanrooms  in the Davis Campus complex (Fig.~\ref{fig:DavisCampus}), near the Yates shaft, at the 4850' level ($\sim$4260 m.w.e.~\cite{Heise2013}) of the mine. These rooms consist of a detector room, a general purpose lab space, and a machine shop. Figure~\ref{fig:LabLayout} shows a picture of the Detector Room at SURF. This room will house the \DEM\ and has dimensions of 32'$\times$40'.  Next to the Detector Room is the Machine Shop. Here, all the copper and NXT-85 parts for the \DEM\ are fabricated, thus further reducing the UGEFCu's surface exposure to cosmic rays. The Machine Shop is approximately 1000 ft$^2$ and includes: two lathes, two mills, an oven, a wire electric discharge machine, a press, a drill press and a laser engraver. The final room is a general purpose lab and is used for testing the detectors prior to their installation into the \DEM. The room is approximately 550 ft$^2$ and was originally designed for electroforming activities; hence its formal name is the Electroforming Room.

\begin{figure}[!htbp]
\begin{center}
\includegraphics[width=7cm]{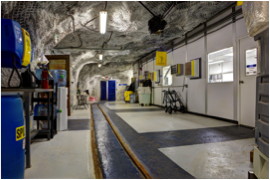}
\caption{A photograph of the TCR used for electroforming on the 4850' level at SURF.}
\label{fig:TCRpic}
\end{center}
\end{figure}

The final \MJ\ laboratory is the Temporary Cleanroom (TCR) (not pictured in Fig.~\ref{fig:LabLayout}), which sits on the same level as the Davis Campus and is approximately 1 km away. It consists of a cleanroom building, shown in Fig.~\ref{fig:TCRpic}, that contains 10 electroforming baths and a small annex for changing into cleanroom garb. The total area of the building is 12'$\times$40' with the annex room consuming 8'$\times$12' of that area. The TCR was required, prior to beneficial occupancy of the Davis Campus, for initiating the slow process of electroforming copper in order for parts to be ready on time for assembly of the \DEM.

\section{The Background Model and the \MJDEMbf\ Sensitivity}
\label{Sec:Background}
The projected background in the \DEM\ is significantly improved over previous generation experiments. This reduction is a result of fielding the detectors in large arrays that share a cryostat and minimizing the amount of interstitial material. Further background suppression is achieved through the aggressive reduction of radioactive impurities in construction materials and minimization of exposure to cosmic rays.  \MJ\ will also make use of event signatures to reject backgrounds that do appear, including pulse-shape characteristics, detector hit granularity, cosmic ray veto tags, and single-site time correlations. In this section we describe these aspects of the \MJ\ \DEM\ design and their impact on the projected backgrounds and physics sensitivity.

\subsection{Pure Materials}

The production process for enriched germanium detectors (enrichment, zone refining, and crystal growth) efficiently removes natural radioactive impurities from the bulk germanium. The cosmogenic activation isotopes, \nuc{60}{Co} and \nuc{68}{Ge}, are produced in the crystals while they are above ground, but can be sufficiently reduced by minimizing the time to deployment underground and by the use of passive shielding during transport and storage. 

For the main structural material in the innermost region of the apparatus, we choose copper for its lack of naturally occurring radioactive isotopes and its excellent physical properties. By starting with the cleanest copper stock we have identified and then electroforming it underground to eliminate primordial radioactivity and cosmogenically-produced \nuc{60}{Co}, we have achieved several orders-of-magnitude background reduction over commercial alternatives. Electroformed copper will also be employed for the innermost passive, high-Z shield. Commercial copper stock is clean enough for use as the next layer of shielding. For all uses of copper, we have certified the cleanliness of samples via assay. Modern lead is available with sufficient purity for use as the bulk shielding material outside of the copper layers.

Several clean plastics are available for electrical and thermal insulation. For the detector supports we use a pure Polytetrafluoroethylene (PTFE), DuPont\texttrademark ~Teflon\textregistered ~NXT-85. Thin layers of low-radioactivity parylene will be used as a coating on copper threads to prevent galling, and for the cryostat seal. For the few weight-bearing plastic components requiring higher rigidity, we have sourced pure stocks of PEEK\textregistered ~(polyether ether ketone), produced by Victrex\textregistered, and Vespel\textregistered, produced by DuPont\texttrademark.

The front-end electronics are also designed to be low-mass and ultra-low background because they must be located in the interior of the array adjacent to the detectors in order to maintain signal fidelity. The circuit board is fabricated by sputtering thin traces of pure gold and titanium on a silica wafer, upon which a bare FET is mounted using silver epoxy. A $\sim$G$\Omega$-level feedback resistance is provided by depositing intrinsically pure amorphous Ge. Detector contact is made via an electroformed copper pin with beads of low-background tin at either end. An electroformed-copper spring provides the contact force. Our signal and high-voltage cables are extremely low-mass  miniature coaxial cable. We have worked with the vendor to cleanly fabricate the final product using pure stock that we provide for the conductor, insulation, and shield. Cable connectors within the cryostat are made from electroformed copper, PTFE, and the same silica circuit boards used for the front-end electronics.

The high material purities required for the \MJ\ \DEM\ necessitated the development of improved assay capabilities. These capabilities are needed not just to establish that the required purities can be achieved, but to also monitor construction processes to verify that cleanliness is maintained. We rely primarily on three assay methods: $\gamma$-ray counting, inductively-coupled plasma mass spectrometry (ICP-MS), and neutron activation analysis (NAA).

\subsection{Background Rejection}
\label{Sec:BackgroundRejection}

One key advantage of HPGe detectors is their inherent excellent energy resolution. Background rejection in our \ppc\ detectors is significant due to, not only the energy resolution, but also array granularity (inter-detector coincidences), pulse-shape discrimination, and event time correlation. These techniques rely on the differentiation of the spatial and temporal distributions of \BBz-decay events from most background events. Background signals from radioactive decay often include a $\beta$ and/or one or more $\gamma$ rays. Such $\gamma$ rays frequently undergo multiple scatters over several centimeters. Since \BBz-decay energy deposition typically occurs within a small volume ($\approx$1 mm$^3$), it is a single site energy deposit. The two different topologies can be separated by PSA (see Sec.~\ref{Sec:PPCTech}). Rejection of decay-product series using single-site time-correlation analysis techniques is also possible in these ultra-low event rate experiments. For example,  \nuc{68}{Ga} $\beta^+$ decays can only result in background to \BBz\ decay if one of the annihilation $\gamma$ rays interacts in the same crystal that contains the $\beta^+$. Hence it is always a multiple-site energy deposit and we reject much of this background through PSA. In addition, however, \nuc{68}{Ga} decays are preceded by the electron capture decay of the parent \nuc{68}{Ge}. The low threshold of the \ppc\ detectors permits additional rejection by a time-correlation cut with the \nuc{68}{Ge} 10-keV K and 1-keV L x rays. 

\begin{figure*}[!htbp]
\includegraphics[width=15cm]{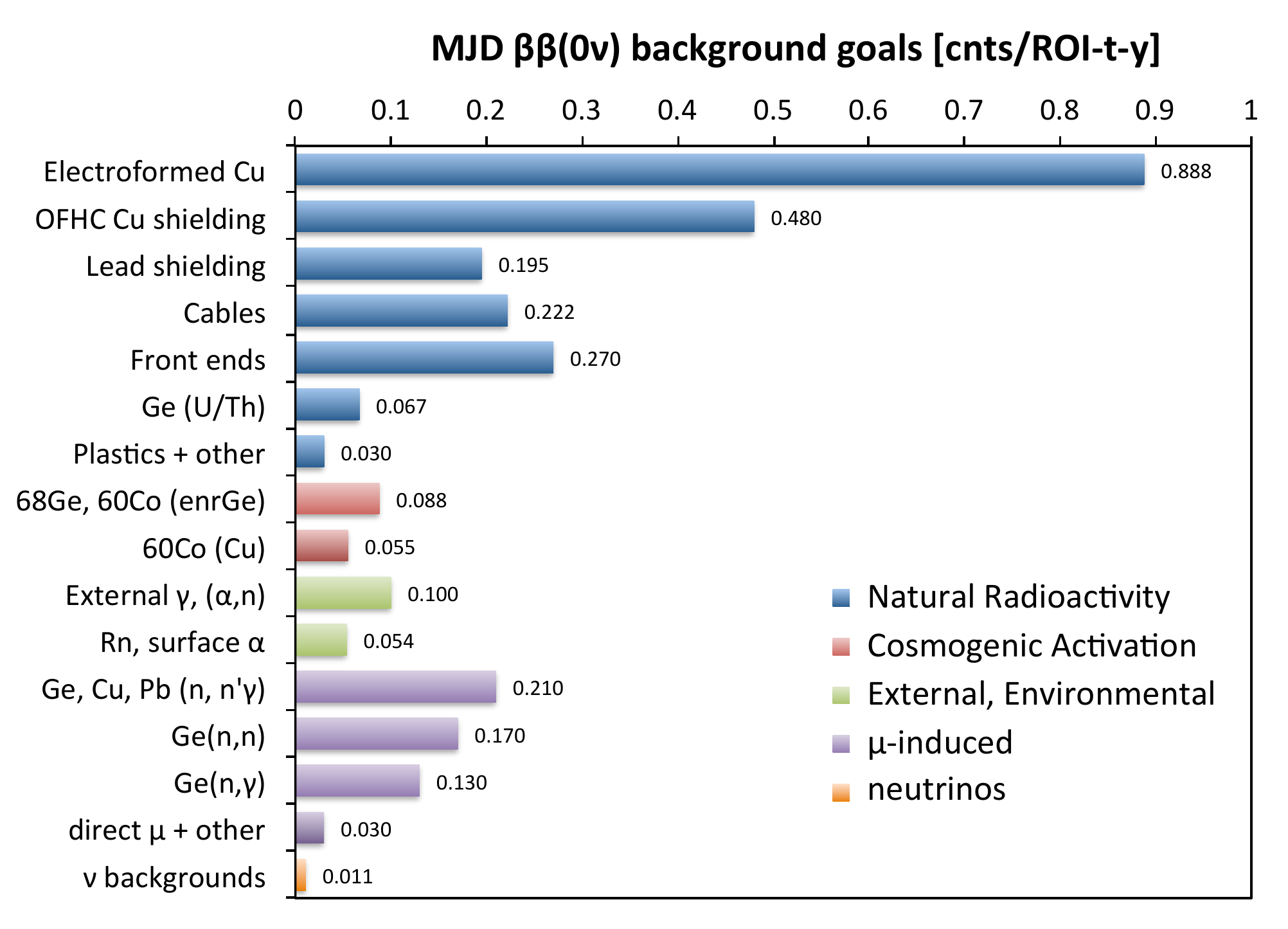}
\caption{Estimated background contributions for the \BBz\ decay search. Backgrounds from natural radioactivity are shown in blue. Cosmogenic activation products are colored red. Green denotes backgrounds from the environment or those introduced during detector assembly. Purple is for $\mu$-induced backgrounds at depth. An upper limit on the negligible background from atmospheric and other neutrinos is shown in orange. The contributions sum to 3.0 \cpRty\ in the \MJ\ \DEM.}
\label{fig:BGsummary}
\end{figure*}

\subsection{Monte Carlo Simulations}
\label{Sec:Simulations}

The \MJ\ and \Gerda\ collaborations have jointly developed a simulation software framework, \MaGe, that is based on the
\GF\ ~\cite{Agostinelli2003250,All06} simulation toolkit. \MaGe\ is used to simulate the response of  
ultra-low radioactive background detectors to ionizing radiation. The development of a common simulation framework used by both collaborations reduces duplication of effort, eases comparison between simulated data and experiment, and simplifies the addition of new simulated detector geometries. The \MaGe\ package is described in more detail in Ref.~\cite{bos11}.

\MaGe\ has interfaces with numerous external packages including software that simulates charge pulse generation in HPGe detectors.  The \MaGe\ framework contains the geometry models of common detector objects, experimental prototypes,  
test stands, and the \DEM\ itself. 
It also implements customized event generators, \GF\ physics lists, and output 
formats. All of these features are available as class libraries that are typically compiled into a single executable. The user selects the particular experimental setup implementation at run-time via macros.  In the prototyping phase, the simulation was used as a virtual test stand guiding detector design, allowing an estimate of the effectiveness of proposed background reduction techniques, and providing an estimate
of the experimental sensitivity. During  operation, \MaGe\ will be used to simulate and characterize backgrounds to determine the ultimate sensitivity of the experiment. It will also provide probability distribution functions (PDFs) for likelihood-based signal extraction analyses. 


\MJ\ has developed a detailed geometry of the \DEM\ in \MaGe\ that consists of more than $3,800$ components. The collaboration has performed detailed simulations of $60,000$ background contributions from different isotopes in these components that required 60 kCPU-hrs on different Linux clusters. The simulations include estimates of the rejection efficiencies from the analysis cuts discussed in Sec.~\ref{Sec:BackgroundRejection}. The PSA cut efficiencies, in particular, are estimated using a heuristic calculation in which multiple interactions in a detector are examined for their relative drift time using isochrone maps such as that depicted in Fig.~\ref{fig:DriftCalc}. That information is combined with the energies of the interactions to determine whether the PSA algorithm would be capable of rejecting the event. The summary of the background expectation after all cuts is given in Fig.~\ref{fig:BGsummary}. These simulation results were used for the current background estimates and will be used in future analyses of data. 

The collaboration has also developed an automated validation suite that thoroughly tests all the critical physics process that are being simulated by \MaGe\ and \GF\ against validated experimental data. This suite is run every time there is a major update to \MaGe\ or \GF\ to verify that the critical physics processes are not altered between versions.

\subsection{Predicted Sensitivity of the \MJDEMbf}
\begin{figure*}[!htbp]
\begin{center}
\includegraphics[width=15cm]{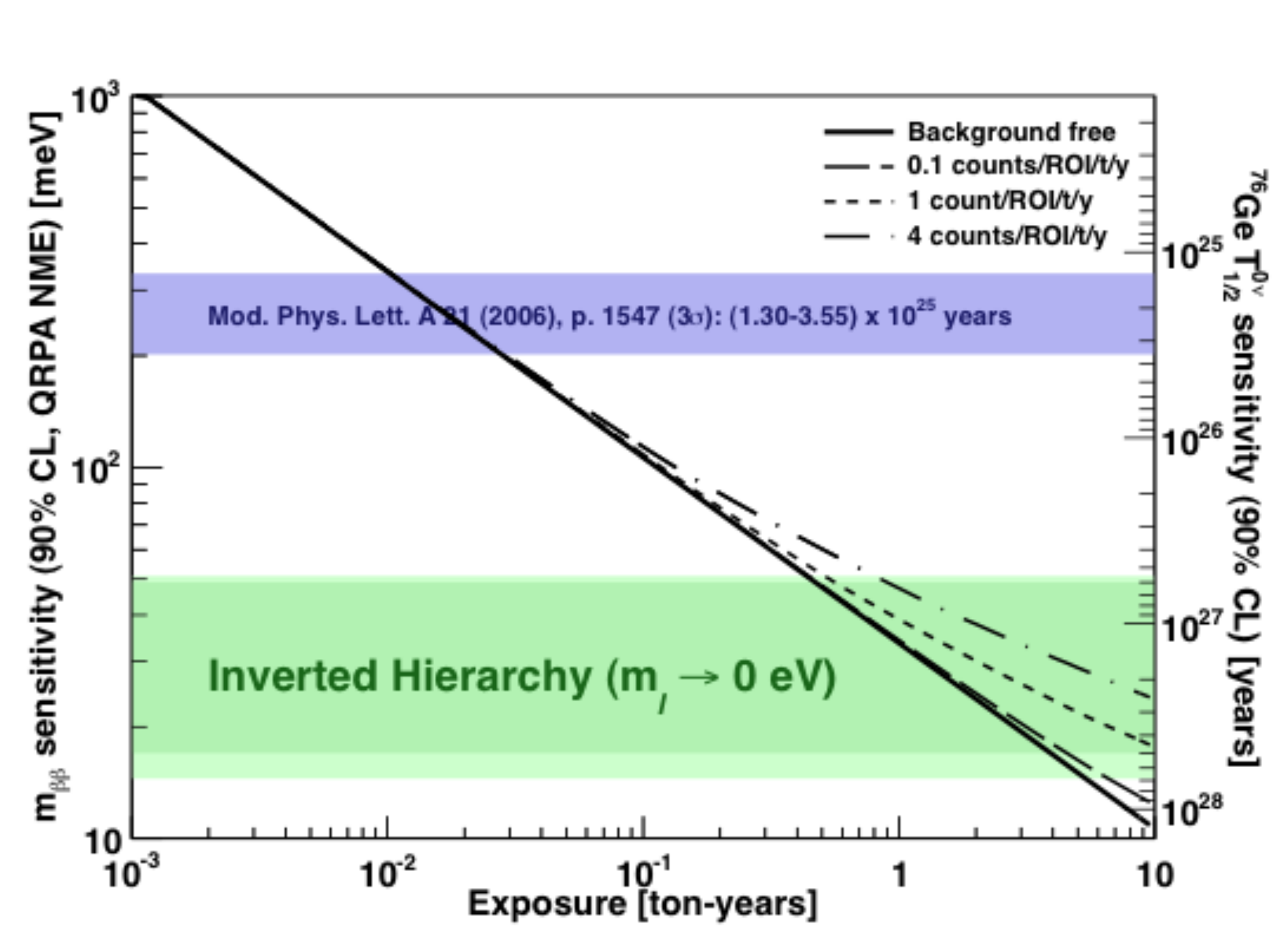}
\caption{90\% C.L. sensitivity as a function of exposure for \BBz-decay searches in \nuc{76}{Ge} under different background scenarios. Matrix elements from Ref.~\cite{Sim09} were used to convert half-life to neutrino mass.  The blue band shows the region where a signal would be detected if the recent claim~\cite{kla06} is correct.}
\label{fig:Sensitivity}
\end{center}
\end{figure*}

The sensitivity of a neutrinoless double-beta decay search increases with the exposure of the experiment, but ultimately depends on the achieved background level. This relationship is illustrated for the \DEM\ in Fig.~\ref{fig:Sensitivity}, in which we have used the Feldman-Cousins~\cite{Feldman1998} definition of sensitivity in order to transition smoothly between the background-free and background-dominated regimes. The background expectation for the \DEM\ is \threecpRty in a 4-keV energy window at the \BBz\ decay endpoint energy. A minimum exposure of about 30 kg-y is required to test the recent claim of an observation of \BBz\ decay~\cite{kla06}.

\section{Status, Prospects, and Conclusions}
\subsection{Status of the \MJDEMbf\ Project}
The \MJ\ Collaboration obtained beneficial occupancy of its Davis Campus underground laboratories in May 2012. At the time of this writing in June 2013, the collaboration has completed outfitting the labs, established cleanliness (the detector room is typically better than 500 particles/ft$^3$ of diameter 0.5 $\mu$m or smaller), and is proceeding with the construction and assembly of the array. The underground electroforming laboratories, which started operation in the summer of 2011, have now produced more than 75\% of the required copper.  The 42.5~kg of 86\% enriched $^{76}$Ge has been reduced from GeO$_2$ and refined to electronic grade Ge with a yield of 98\%. Ten enriched \ppc\ detectors with a total mass of 9.5 kg have been produced by ORTEC, with eight now underground to SURF.  A prototype cryostat, the same as the ultra-clean cryostats, but fabricated from commercial copper,  has been assembled and operated with its associated vacuum system. Two strings of natural Ge detectors have been built in the glove boxes and are undergoing testing before they are installed in the Prototype Cryostat.  Cryostat 1 has been machined from the UGEFCu.  Samples obtained from  all materials being used in the \DEM\ are being assayed. Slow control systems and their associated sensors are in continuous operation in all UG laboratories.  Data acquisition systems for detector acceptance testing, string testing, and the main array are operational.

The Prototype Cryostat will be commissioned in the summer of 2013.  Cryostat 1, which will contain seven strings of both enriched and natural Ge detectors, is scheduled to be completed in late 2013. Cryostat 2, which is expected to  contain all enriched detectors, is expected to be assembled in 2014. The full array should be in operation in 2015.  The \DEM\ will be operated for about 3 years in order to collect $\sim$100 kg-years of exposure.  

\subsection{The Future Large-Scale Experiment}
 
The \MJ\ and GERDA collaborations are working together to prepare for a single tonne-scale \nuc{76}{Ge} experiment that will combine the best technical features of both experiments.  The results of the two experiments will be used to determine the best path forward.

The present generation of experiments will likely produce results with limits on \mee\ below about 100 meV. The next generation will strive to cover the inverted hierarchy region of the effective Majorana neutrino mass. To accomplish this will require \mee\ sensitivity down to about 20 meV. Such small neutrino masses would indicate a half-life longer than $10^{27}$ y. As seen in Fig.~\ref{fig:Sensitivity}, to observe such a long half-life, one will need a tonne or more of isotope and backgrounds below \onecpRty.

\section*{Acknowledgments}
We acknowledge support from the Office of Nuclear Physics in the DOE Office of Science under grant 
numbers DE-AC02-05CH11231, DE-FG02-97ER41041, DE-FG02-97ER41033, DE-FG02-97ER4104, 
DE-FG02-97ER41042, DE-SCOO05054, DE-FG02-10ER41715, and DE-FG02-97ER41020. We acknowledge support 
from the Particle and Nuclear Astrophysics Program of the National Science Foundation through grant 
numbers PHY-0919270, PHY-1003940, 0855314, PHY-1202950, MRI 0923142 and 1003399. We gratefully acknowledge support from the 
Russian Foundation for Basic Research, grant No. 12-02-12112. We gratefully acknowledge the support of the U.S. Department 
of Energy through the LANL/LDRD Program.

\section{References}
\addcontentsline{toc}{chapter}{Bibliography}
\bibliographystyle{iopart-num.bst}
\bibliography{DoubleBetaDecay.bbl}

\end{document}